\documentclass[modern]{aastex61}
\shorttitle{Predicting Solar Flares using Vector Magnetic Data}
\shortauthors{Liu et al.}

\def\mathbi#1{\textbf{\em #1}}
\newcommand{\sm}{$\sim$}

\usepackage[T1]{fontenc}
\usepackage{textcomp}

\begin{document}
\title{PREDICTING SOLAR FLARES USING SDO/HMI VECTOR MAGNETIC DATA PRODUCT AND RANDOM FOREST ALGORITHM}

\author[0000-0002-6178-7471]{Chang Liu}
\affiliation{Space Weather Research Laboratory, New Jersey Institute of Technology, University Heights, Newark, NJ 07102-1982, USA; chang.liu@njit.edu, na.deng@njit.edu, haimin.wang@njit.edu}
\affiliation{Big Bear Solar Observatory, New Jersey Institute of Technology, 40386 North Shore Lane, Big Bear City, CA 92314-9672, USA}
\affiliation{Center for Solar-Terrestrial Research, New Jersey Institute of Technology, University Heights, Newark, NJ 07102-1982, USA}

\author{Na Deng}
\affiliation{Space Weather Research Laboratory, New Jersey Institute of Technology, University Heights, Newark, NJ 07102-1982, USA; chang.liu@njit.edu, na.deng@njit.edu, haimin.wang@njit.edu}
\affiliation{Big Bear Solar Observatory, New Jersey Institute of Technology, 40386 North Shore Lane, Big Bear City, CA 92314-9672, USA}
\affiliation{Center for Solar-Terrestrial Research, New Jersey Institute of Technology, University Heights, Newark, NJ 07102-1982, USA}

\author{Jason T. L. Wang}
\affiliation{Department of Computer Science, New Jersey Institute of Technology, University Heights, Newark, NJ 07102-1982, USA; jason.t.wang@njit.edu}

\author{Haimin Wang}
\affiliation{Space Weather Research Laboratory, New Jersey Institute of Technology, University Heights, Newark, NJ 07102-1982, USA; chang.liu@njit.edu, na.deng@njit.edu, haimin.wang@njit.edu}
\affiliation{Big Bear Solar Observatory, New Jersey Institute of Technology, 40386 North Shore Lane, Big Bear City, CA 92314-9672, USA}
\affiliation{Center for Solar-Terrestrial Research, New Jersey Institute of Technology, University Heights, Newark, NJ 07102-1982, USA}

\begin{abstract}
Adverse space weather effects can often be traced to solar flares, prediction of which has drawn significant research interests. The Helioseismic and Magnetic Imager (HMI) produces full-disk vector magnetograms with continuous high cadence, while flare prediction efforts utilizing this unprecedented data source are still limited. Here we report results of flare prediction using physical parameters provided by the Space-weather HMI Active Region Patches (SHARP) and related data products. We survey X-ray flares occurred from 2010 May to 2016 December, and categorize their source regions into four classes (B, C, M, and X) according to the maximum GOES magnitude of flares they generated. We then retrieve SHARP related parameters for each selected region at the beginning of its flare date to build a database. Finally, we train a machine-learning algorithm, called random forest (RF), to predict the occurrence of a certain class of flares in a given active region within 24 hours, evaluate the classifier performance using the 10-fold cross validation scheme, and characterize the results using standard performance metrics. Compared to previous works, our experiments indicate that using the HMI parameters and RF is a valid method for flare forecasting with fairly reasonable prediction performance. To our knowledge, this is the first time that RF is used to make multi-class predictions of solar flares. We also find that the total unsigned quantities of vertical current, current helicity, and flux near polarity inversion line are among the most important parameters for classifying flaring regions into different classes.
\end{abstract}

\section{INTRODUCTION}
Solar flares and the often associated coronal mass ejections (CMEs) can severely impact the near-Earth space environment, causing geomagnetic and particle disturbances with potentially deleterious technological and societal consequences \citep{daglis04}. Building the space-weather readiness merits substantial efforts on several fronts, including research, forecast, and mitigation plan, as recognized by the recently released U.S. National Space Weather Strategy.

Observational and theoretical research has suggested that flares and CMEs are powered by magnetic free energy (difference between potential and non-potential magnetic energy) accumulated in the corona and rapidly released by magnetic reconnection \citep{priest02}. This build-up process of coronal free energy is essentially governed by the structural evolution of magnetic field on the photosphere, where the plasma dominates and on which the coronal field is anchored. Thus, although direct measurement of the weak coronal magnetic field is challenging, structure and evolution of the photospheric magnetic field, which can be observed and measured, may provide critical clues to the energy accumulation and triggering mechanisms of flares/CMEs \citep[for a review see e.g.,][]{wang15}. The static and evolving photospheric magnetic structural properties of active regions (ARs) can be characterized by a variety of parameters, such as size and complexity (described by e.g., sunspot classification schemes), vertical electric currents, surface magnetic free energy, unsigned magnetic flux, integrated Lorentz force, magnetic shear and gradient, magnetic energy dissipation, and magnetic helicity injection.

Although substantial efforts have been invested, details regarding the physical relationship between the flare productivity and non-potentiality of ARs as reflected by the above parameters are still far from being fully understood. Nevertheless, the fundamental magnetic coupling between the photosphere and the corona has motivated the use of photospheric field parameters for predicting flares, not based on physical flare models but on various approaches of statistics and machine learning \citep[see][and references therein]{bloomfield12,barnes16}. In particular, machine learning is a subfield of computer science that enables algorithms to learn from the input (training) data and make data-driven predictions. It automates analytical model building, thus allows hidden insights to be discovered from data. Most of previous studies used parameters derived from the line-of-sight (LOS) component of the photospheric magnetic field, and produced probability outputs for the occurrence of a certain magnitude flare in a time period. For examples, \citet{gallagher02b} and \citet{bloomfield12} used the McIntosh sunspot classification system and Poisson statistics to estimate probabilities of an AR to produce flares with different magnitude in 24 hours. \citet{song09} adopted three LOS magnetic parameters and employed the ordinal logistic regression (OLR) method to yield one-day flare probabilities. As pointed out by \citet{bloomfield12}, the predicted probabilities may need to be converted into a yes-or-no forecast before the result can be practically interpreted as ``flare imminent'' or ``flare quiet''. Recognizing that such a conversion in \citet{song09} was accomplished by manually chosen threshold values, \citet{yuan10} enhanced their results by feeding the obtained probabilities into multiple binary classifiers each called a support vector machine (SVM) to obtain a definite true or false prediction of flares with different classes.

Compared to the LOS field, the full vector data supplies more information about the photospheric magnetic field structure that may warrant a better prediction performance; however, efforts on flare forecast using vector field parameters were restricted, mainly due to the availability limitation imposed by ground-based vector magnetic field observation. \citet{leka03} first used a small sample of vector magnetograms from the Mees Solar Observatory and applied a discriminant analysis to distinguish flare-producing and flare-quiet ARs within few hours. Subsequent studies were also made on extending to a larger number of samples and a prediction time window of 24 hours \citep{leka07}, and on producing probability forecast \citep{barnes07}. It is notable that since 2010 May, the Helioseismic and Magnetic Imager (HMI; \citealt{schou12}) on board the Solar Dynamics Observatory (SDO; \citealt{pesnell12}) has been producing unprecedented photospheric vector magnetograms with continuous high-cadence (normally 12 minutes), full-disk coverage. One of the key science questions for the SDO mission is ``When will activity occur, and is it possible to make accurate and reliable forecasts of space weather and climate?'' Using four years SDO/HMI vector field data since its launch, \citet{bobra15} calculated a number of magnetic parameters for each AR. The authors chose 13 parameters, most of which can be only derived from vector data, and achieved good predictive performance with a SVM method for flares greater than M1.0 class, as defined by the peak 1--8~\AA\ flux measured by a Geostationary Operational Environmental Satellite (GOES). In a recent work of \citet{nishizuka17}, the authors applied a number of machine learning algorithms to HMI vector data and also ultraviolet brightenings, and developed prediction models for $\geqslant$M and X-class flares with high performance. Certainly, more flare forecast studies using HMI data are desired in order to fully explore their prediction capability. 

In this paper, we attempt to use SDO/HMI vector data to forecast the maximum magnitude of flares in terms of GOES classes (i.e., B, C, M, and X) that would occur in a given AR within 24 hours, with a machine learning algorithm called random forest \citep[RF;][]{breiman01}, which is based on an ensemble of CART-like decision tress \citep{breiman84}. RF can very well handle high dimensional feature space and do not expect linear features as compared to the OLR method. Also, unlike SVM that is fundamentally a binary classifier, RF is an inherent multi-class classifier. Other advantages of RF include being a highly accurate learning algorithm, no need to pre-process data, and resistance to over-training\footnote{\url{https://www.stat.berkeley.edu/~breiman/RandomForests/}\label{note1}}. The RF has been successfully used in science informatics to perform, for example, biological data analysis \citep[e.g.,][]{laing12}. There has also been many successful applications of RF in astronomy, such as identifying quasars \citep{breiman03}, estimating photometric redshifts \citep{carliles10}, searching for supernova \citep{bailey07} and gravitational waves \citep{hodge14}, and auto classification of astrophysical sources \citep{farrell15}. One method used by \citet{nishizuka17} is the extremely randomized trees (ERT), which is similar to RF (see more discussions in Section~\ref{method}). To our knowledge, this is the first time that RF is used to make multi-class predictions of solar flares.

The plan of the paper is as follows. In Section~\ref{data}, we describe the predictive parameters and sample selection, and also study the general properties of the samples. In Section~\ref{method}, we introduce the RF algorithm and schemes for result validation and performance evaluation. Major results are presented and discussed in Section~\ref{result}, and a summary is given in Section~\ref{summary}.

\section{FLARE PREDICTIVE PARAMETERS AND AR SAMPLES} \label{data}
Near the end of 2012, the SDO/HMI team began to release a data product called Space-weather HMI Active Region Patches (SHARP; \citealt{bobra14}), with a main goal of facilitating AR events forecasting. These derivative data, available as the \verb|hmi.sharp| data series from the Joint Science Operations Center (JSOC)\footnote{\url{http://jsoc.stanford.edu/}}, encompass automatically identified and tracked ARs in map patches and provide many magnetic measurements and derived physical parameters via map quantities and keywords. In mid 2014, a separate data series \verb|cgem.Lorentz| was produced based on SHARP data to include estimations of integrated Lorentz forces \citep{sun14}, which can help diagnose dynamic processes of ARs \citep{fisher12}. In total, 25 parameters characterizing AR magnetic field properties are calculated for selected ARs and are contained in the above SHARP related data products. Using a univariate feature selection algorithm, \citet{bobra15} scored these parameters and suggested the use of the top 13 (listed in Table~\ref{sample_overview} in the order of their rankings) as predictors for flaring activity.

\begin{table}[!t]
\begin{center}
\caption{Overview of AR Samples using SDO/HMI Magnetic Parameters\label{sample_overview}}
\setlength{\tabcolsep}{1.pt}
\scriptsize
\begin{tabular}{lccccccc}
\hline\noalign{\smallskip}
SHARP &  & & RF & B Class & C Class & M Class & X Class \\
Keyword$^{a}$ & Formula & Unit & Importance$^{b}$ &($n=128$) & ($n=552$) & ($n=142$) & ($n=23$)\\
\hline\noalign{\smallskip}
TOTUSJH & ${H_{c_{\rm total}}} \propto \sum |B_z \cdot J_z|$ & 10$^2$~G$^2$~m$^{-1}$ & 37.4 &4.8$\pm$3.1 & 13.9$\pm$9.9 & 27.7$\pm$18.2 & 58.3$\pm$40.0 \\
TOTBSQ  & $F \propto \sum B^{2} $ & 10$^{10}$~G$^2$ & 17.9 & 1.0$\pm$0.9 & 2.6$\pm$1.9 & 4.6$\pm$3.0 & 10.7$\pm$8.6 \\
TOTPOT & $ \rho_{\rm tot} \propto  \sum \left( \mathbi{B}^{\rm Obs} - \mathbi{B}^{\rm Pot} \right)^2 dA $ & 10$^{23}$~ergs~cm$^{-3}$ & 21.1 & 1.0$\pm$1.4 & 2.7$\pm$2.7 & 6.7$\pm$5.7 & 19.6$\pm$18.0\\
TOTUSJZ & ${J_{z_{\rm total}}} =  \sum |J_{z}|dA$ & 10$^{12}$~A & 50.6 & 9.5$\pm$6.4 & 30.3$\pm$21.4 & 53.9$\pm$30.9 & 110.0$\pm$73.4 \\
ABSNJZH & ${H_{c_{\rm abs}}} \propto \left| \sum B_z \cdot J_z \right|$ & 10~G$^2$~m$^{-1}$ & 19.9 & 6.1$\pm$7.0 & 14.3$\pm$17.0 & 39.2$\pm$43.8 & 91.2$\pm$63.6 \\
SAVNCPP & $J_{z_{\rm sum}} \propto \Big\vert \displaystyle\sum\limits^{B{_z^+}} J{_z}dA \Big\vert + \Big\vert \displaystyle\sum\limits^{B{_z^-}} J{_z}dA \Big\vert $ & 10$^{12}$~A & 24.6 & 2.7$\pm$2.7 & 6.5$\pm$6.4 & 15.8$\pm$14.6 & 33.1$\pm$24.0 \\
USFLUX & $\Phi = \sum|B_{z}|dA$ & 10$^{21}$~Mx & 14.2 & 7.1$\pm$5.5 & 19.9$\pm$14.7 & 33.7$\pm$21.0 & 72.2$\pm$54.2 \\
AREA\_ACR & Area $ = \sum$ Pixels & 10$^{2}$ pixels & 23.7 & 3.0$\pm$2.4 & 8.2$\pm$6.1 & 13.3$\pm$7.7 & 29.2$\pm$22.3 \\
TOTFZ  & $F_{z}  \propto \sum (B_{x}^{2} + B_{y}^{2} - B_{z}^{2}) dA$ & $-10^{23}$~dyne & 13.9 & 1.2$\pm$1.3 & 2.7$\pm$2.7 & 3.9$\pm$3.7 & 6.1$\pm$6.2 \\
MEANPOT & $ \overline{\rho} \propto \frac{1}{N} \sum \left( \mathbi{B}^{\rm Obs} - \mathbi{B}^{\rm Pot} \right)^2 $ & $10^3$~ergs~cm$^{-3}$ & 19.8 & 6.5$\pm$5.8 & 5.9$\pm$3.7 & 8.9$\pm$4.2 & 12.1$\pm$4.2 \\
R\_VALUE & $\Phi = \sum|B_{{\rm LOS}}|dA$ within $R$ mask & Mx & 31.4 & 3.2$\pm$0.7 & 3.8$\pm$0.6 & 4.4$\pm$0.5 & 4.9$\pm$0.4 \\
EPSZ & $\delta F_{z}  \propto \frac{\sum (B_{x}^{2} + B_{y}^{2} - B_{z}^{2})}{ \sum B^{2}}$ & $-10^{-1}$ & 15.4 & 2.1$\pm$1.3 & 2.0$\pm$1.3 & 1.7$\pm$1.2 & 1.2$\pm$1.1 \\
SHRGT45 & Area with shear $>45^\circ$ / Total Area & -- & 12.7 &  0.23$\pm$0.17 & 0.27$\pm$0.14 & 0.34$\pm$0.13 & 0.40$\pm$0.11\\
\noalign{\smallskip}\hline
\end{tabular}
\end{center}
\tablecomments{(a) These 13 SDO/HMI magnetic parameters are ordered according to their rankings by univariate Fisher scores as evaluated by \citet{bobra15}. Their values of the 845 AR samples are extracted from HMI SHARP related data products. The values shown represent mean plus/minus 1$\sigma$ calculated for each of the four AR class samples. The meaning of each parameter is as follows: total unsigned current helicity (TOTUSJH), total magnitude of Lorentz force (TOTBSQ), total photospheric magnetic free energy density (TOTPOT), total unsigned vertical current (TOTUSJZ), absolute value of the net current helicity (ABSNJZH), sum of the modulus of the net current per polarity (SAVNCPP), total unsigned flux (USFLUX), area of strong field pixels in the active region (AREA\_ACR), sum of $z$-component of Lorentz force (TOTFZ), mean photospheric magnetic free energy (MEANPOT), sum of flux near polarity inversion line (R\_VALUE), sum of $z$-component of normalized Lorentz force (EPSZ), and fraction of area with shear $> 45^\circ$ (SHRGT45). (b) The RF importance values are of the type of mean decrease Gini.}
\end{table}

The machine learning technique relies on training samples. For this study, we surveyed flares that occurred in a $\sim$6.5-year period (from the SDO launch time in 2010 May to 2016 December) covering the main peak of the solar cycle 24, using the GOES X-ray flare catalogs\footnote{\url{https://www.ngdc.noaa.gov/nndc/struts/results?t=102827&s=25&d=8,230,9} Note that at the time of the present work, the 2011 flare catalog has no associated AR location information. We thus recreated it using the IDL procedure \texttt{xraydatareports.pro} by Dr. William Denig.} prepared by the National Centers for Environment Information (NCEI; formerly the National Geophysical Data Center). These catalogs are constructed by merging the monthly GOES X-ray flare listings (providing information of flare time and peak magnitude in 1--8~\AA\ soft X-ray flux, source AR, etc.) with the associated H-alpha flare listings (providing information of flare time, location, source AR, etc.) from the USAF Solar Observing Optical Network \citep{denig12}. We then built a database of flare-producing ARs in the following way. (1) We used a four-class (i.e., B, C, M, and X) AR classification scheme \citep[e.g.,][]{song09,yuan10}, which is determined based on the maximum GOES-class flare(s) an AR ever produces. This means that an AR classified into a certain class produces at least one flare with such GOES class but no flares with higher GOES class. Note that the B class is the lowest flare class listed in the NCEI catalogs. (2) We only selected C-, M-, and X-class ARs with records of flares that have identified locations in the NCEI flare catalogs and occur within about $\pm70^{\circ}$ of the central meridian (for the optimum vector field data quality; see \citealt{bobra14}). To maximize the number of B-class AR samples, we considered all B-class ARs, and for those with no location information in the NCEI catalogs, we manually checked with solarmonitor.org. (3) We made sure that valid values of all the 13 magnetic parameters of the selected AR samples at the beginning of the flare day are available from SDO/HMI data products (see discussions below). The measurement must also produce reliable Stokes vectors, with the value of the QUALITY keyword smaller than 65,536 \citep{bobra14}. (4) If an AR produces multiple flares with the same GOES class on the same day, only one sample corresponding to the last such flare was recorded; however, if these flares occur on different dates, the records were treated as different valid samples. (5) We caution that the NCEI catalogs may contain errors on the association between flares and their source ARs. For example, the 2014 October 25 X1.0 flare peaked at 17:08 UT was located at a major flaring region NOAA AR 12192, but a small quiet region AR 12196 was assigned to this flare in the catalog. This would negatively affect the training of the machine learning algorithm, because the magnetic parameter values of the flare-quiet AR 12196 differ significantly from those of the true X-class flaring regions. Several entries with such obvious inconsistencies were manually corrected. In total, we collected 845 samples including 23 X-class, 142 M-class, 552 C-class, and 128 B-class ARs (see Table~\ref{samples} in appendix).

\begin{figure}[!t]
\epsscale{1.2}
\plotone{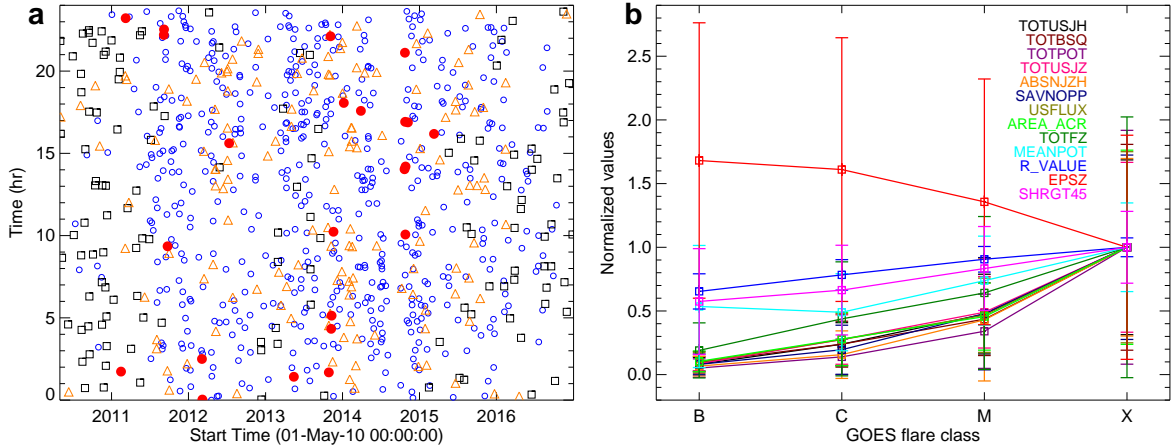}
\caption{Properties of 845 flaring AR samples. (a) Distribution of flare start time in GOES 1--8~\AA. B-, C-, M-, and X-class AR samples are denoted as black square, blue circle, orange triangle, and red filled circle, respectively. (b) Mean value (with 1$\sigma$ error bar) of SDO/HMI magnetic parameter normalized to that of the X-class vs. the class of AR samples; also see Table~\ref{sample_overview}. \label{f1} }
\end{figure}

In Figure~\ref{f1}(a) we plot the distribution of the flare start time in GOES 1--8~\AA\ of all the AR samples, which shows a semi-homogeneous spread within a 24 hr time period. Therefore, in this study we retrieve the values of the 13 HMI predictive parameters for each AR in our database\footnote{The complete data set and also the source code for the main experiment in Section~\ref{experiments} can be found at \url{https://web.njit.edu/~cl45/Fpredict/}\label{note2}}
 at the time of the beginning of its flare date (mostly at 00:12~UT), using the routine \verb|ssw_jsoc_time2data| from the SolarSoft package. Since flares originated from our sample ARs can occur anytime within a one day period as shown above, choosing AR parameter values measured at the beginning of the flare date for all samples is in accordance with our objective of predicting flares within 24 hours.

\begin{table}[!t]
\begin{center}
\caption{Spearman Correlation Coefficients $\rho$ among AR Class and 13 SDO/HMI Magnetic Parameters \label{spearman}}
\setlength{\tabcolsep}{3.3pt}
\scriptsize
\begin{tabular}{l|ccccccccccccc}
\hline\noalign{\smallskip}
 & TOT & TOT & TOT & TOT & ABS & SAVN & US & AREA & TOT & MEAN & R\_ & & SHR\\
 & USJH & BSQ & POT & USJZ & NJZH & CPP & FLUX & \_ACR & FZ & POT & VALUE & EPSZ & GT45\\
\hline\noalign{\smallskip}
TOTUSJH & 1 & 0.95 & 0.93 & 0.99 & 0.69 & 0.75 & 0.96 & 0.94 & 0.64 & 0.51 & 0.9 & $-$0.11 & 0.42\\
TOTBSQ  & 0.95 & 1 & 0.94 & 0.93 & 0.61 & 0.67 & 0.97 & 0.93 & 0.7 & 0.53 & 0.84 & $-$0.08 & 0.38 \\ 
TOTPOT & 0.93 & 0.94 & 1 & 0.91 & 0.66 & 0.68 & 0.9 & 0.89 & 0.53 & 0.74 & 0.88 & $-$0.26 & 0.63 \\
TOTUSJZ & 0.99 & 0.93 & 0.91 & 1 & 0.67 & 0.74 & 0.96 & 0.94 & 0.61 & 0.46 & 0.86 & $-$0.13 & 0.42 \\
ABSNJZH & 0.69 & 0.61 & 0.66 & 0.67 & 1 & 0.86 & 0.6 & 0.6 & 0.31 & 0.5 & 0.68 & $-$0.21 & 0.44 \\
SAVNCPP & 0.75 & 0.67 & 0.68 & 0.74 & 0.86 & 1 & 0.67 & 0.69 & 0.38 & 0.43 & 0.72 & $-$0.17 & 0.37\\
USFLUX & 0.96 & 0.97 & 0.9 & 0.96 & 0.6 & 0.67 & 1 & 0.92 & 0.76 & 0.42 & 0.83 & 0.03 & 0.29\\
AREA\_ACR & 0.94 & 0.93 & 0.89 & 0.94 & 0.6 & 0.69 & 0.92 & 1 & 0.55 & 0.48 & 0.79 & $-$0.22 & 0.4 \\
TOTFZ & 0.64 & 0.7 & 0.53 & 0.61 & 0.31 & 0.38 & 0.76 & 0.55 & 1 & 0.01 & 0.53 & 0.61 & $-$0.18 \\
MEANPOT & 0.51 & 0.53 & 0.74 & 0.46 & 0.5 & 0.43 & 0.42 & 0.48 & 0.04 & 1 & 0.63 & $-$0.52 & 0.86 \\
R\_VALUE & 0.9 & 0.84 & 0.88 & 0.86 & 0.68 & 0.72 & 0.83 & 0.79 & 0.53 & 0.63 & 1 & $-$0.16 & 0.51 \\
EPSZ & $-$0.11 & $-$0.08 & $-$0.26 & $-$0.13 & $-$0.21 & $-$0.17 & 0.03 & $-$0.22 & 0.61 & $-$0.52 & $-$0.16 & 1 & $-$0.68 \\
SHRGT45 & 0.42 & 0.38 & 0.63 & 0.42 & 0.44 & 0.37 & 0.29 & 0.4 & -0.18 & 0.86 & 0.51 & $-$0.68 & 1\\
\textbf{AR Class} & 0.74 & 0.68 & 0.67 & 0.74 & 0.59 & 0.64 & 0.68 & 0.68 & 0.4 & 0.39 & 0.71 & -0.16 & 0.33 \\ \hline
\noalign{\smallskip}\hline
\end{tabular}
\end{center}
\end{table}

We further make some exploratory data analyses of our AR samples. First, we compare the magnitude (mean plus/minus 1$\sigma$) of the predictive parameters with AR classes in Table~\ref{sample_overview} and Figure~\ref{f1}(b). It appears that ranging from the B to X class, almost all parameters exhibit a monotonic change of the mean values, with a Pearson correlation coefficient of \sm0.78--0.98. This indicates that ARs producing higher GOES class flares tend to possess larger values of these physical parameters. However, it is also clear that these property values of ARs with different classes could overlap with each other significantly, as the fluctuations in many cases are almost comparable with the means. Thus it is impractical to forecast flares with multiple classes simply based on magnitudes of these quantities. Second, we analyze the correlations $\rho$ among the magnetic parameters and AR classes using Spearman's rank correlation method. The result in Table~\ref{spearman} shows that the AR class is strongly correlated with 9 out of 13 magnetic parameters (with $\rho \gtrsim 0.6$), while it is weakly correlated with the remaining four parameters (with $0.4 \gtrsim |\rho| \gtrsim 0.2$). This implies that these parameters could potentially have good classifying capability of ARs. We also note that there are strong correlations between some parameters, such as TOTUSJH and TOTUSJZ; nevertheless, predicting variables with a high correlation still could be complementary features \citep[e.g.,][]{guyon03}.

\section{METHODOLOGY}\label{method}
We resort to RF, an inherent multi-class classifier, to perform flare prediction. RF is a general term for the random decision forests, an ensemble learning technique mainly for classification and regression tasks \citep{breiman01}. In the training phase, it constructs a collection of decision trees (i.e., tree-like predictive models), each of which is grown based on training records selected using the bagging method (i.e., sampling with replacement), and on randomly selected features when splitting each node. In the testing phase, an unknown sample is evaluated by all decision tress, and the RF outputs the class by majority votes (in the case of  classification) or yields the mean prediction of trees (in the case of regression). Explicitly, the general scheme of RF operates as follows.

\begin{enumerate}
\item Training: Given a training data set $D=d_1,...,d_n$ with responses $R=r_1,...,r_n$, a total of $M$ trees are grown, each of which is built using the following algorithm:

\begin{enumerate}
\item Randomly sample, with replacement, $n$ times to form a training set $(D_m, R_m)$.
\item A decision or regression tree $T_m$ on $(D_m, R_m)$ is trained. At each node split, $f$ variables (predictive parameters) are randomly sampled, and the best split is determined based on information gain or the Gini impurity measure (Gini importance or the mean decrease Gini). By default, $f=\sqrt{F}$ for classification and $f=\frac{F}{3}$ for regression, where $F$ is the total number of variables.
\end{enumerate}

\item Testing/validation: For an unknown sample data record $d'$, forecasting can be made by taking the majority vote for classification decision trees, or can be given by $\frac{1}{M}\sum_{m=1}^{M}T_m(d')$ for regression trees.

\end{enumerate}

\noindent For our flare prediction, we run RF in the classification mode. In the training phase, each input training sample contains the 13 SHARP parameter values of an AR and its corresponding maximum GOES flare class. In the testing/validation phase, the 13 SHARP parameters of an AR sample are used as the model input, and the RF classifier predicts the maximum GOES class of this AR. We note that the ERT method used by \cite{nishizuka17} randomizes, rather than optimizes as what RF does, the splits on trees. A full comparison of these two algorithms is out of the scope of this paper \citep[see][]{geurts06}.

It can be noticed that in our database, there are more C-class AR samples (552) while less X-class samples (23), compared to M-class (142) and B-class (128) samples. In order to alleviate the class imbalance issue that poses a major challenge in machine learning \citep[e.g.,][]{japkowicz02}, we randomly select 142 unique C-class ARs from their total 552 samples to form a complete data set. To avoid any bias, we repeat this random selection 100 times, so we end up with 100 data sets for multi-class classification, each consisting of 128 B-class, 142 C-class, 142 M-class, and 23 X-class AR samples. We also construct another kind of 100 data sets for binary class prediction, so as to facilitate comparisons with previous work. In doing so, we combine the B- and C-class (M- and X-class) to form the B/C (M/X) class, and randomly select (for 100 times) 165 unique B/C-class AR samples to couple with the 165 M/X-class samples.

In evaluating the performance of the RF classifier, we apply the commonly used 10-fold cross validation (CV) method. For each of the data set, we perform a stratified 10-fold partitioning using the function \verb|createFolds| in the \verb|caret| package \citep{kuhn08} in R\footnote{R is a software environment \citep{rsoftware} for statistical computing \citep{james13}, with wide real-world applications in many fields, such as data mining \citep{zhao13}.}. That is, we divide all the samples into 10 groups of nearly equal sizes, which have balanced distributions of AR classes. The prediction function of RF is then trained using nine folds of data, and the one fold left out is used for validation. To account for the random error associated with each 10-fold CV, the CV procedure is repeated 100 times. The average of the total 10$^4$ iterations (100 CVs for each of the 100 data sets for multi-class or binary classification) yields the final results.

To further characterize the prediction results, we consider a confusion matrix, a.k.a. a contingency table for each AR class \verb|k|. The class \verb|k| ARs correctly predicted as class \verb|k| are called true positives (TP), and in case of wrong predictions, they are false negatives (FN). The ARs not in class \verb|k| correctly predicted not as class \verb|k| are true negatives (TN); otherwise they are false positives (FP) if predicted as class \verb|k|. Using these quantities, we compute a various of standard performance metrics. They include recall (a.k.a. sensitive) $= \frac{{\rm TP}}{{\rm TP}+{\rm FN}}$, precision $= \frac{{\rm TP}}{{\rm TP}+{\rm FP}}$, accuracy $= \frac{{\rm TP}+{\rm TN}}{{\rm TP}+{\rm FP}+{\rm TN}+{\rm FN}}$, and the true skill statistics \citep[TSS;][]{hanssen65} defined as

\begin{equation}
{\rm TSS} = \frac{{\rm TP}}{{\rm TP}+{\rm FN}} - \frac{{\rm FP}}{{\rm FP}+{\rm TN}} \ .
\end{equation}

\noindent All these metrics have a value of 1 for perfect forecasts, and are used together for a comprehensive assessment. Because of its unbiasness over class-imbalance ratio \citep{woodcock76}, we follow the suggestion of \citet{bloomfield12} to mainly use the TSS score, which is the recall subtracted by the false alarm rate, when comparing our results with other flare forecasting studies.

\section{RESULTS AND DISCUSSIONS}\label{result}
The RF algorithm was originally written in FORTRAN\textsuperscript{\ref{note1}}. Here we employ the application of RF using the function \verb|randomforest| \citep{liaw02} in R. The function takes in a formula (i.e., combination of predictive parameters) and a training data set, as well as other optional arguments. Two important arguments are \verb|ntree|, which defines the number of trees to be built for the ensemble, and \verb|mtry|, which is the number of variables sampled at each node for splitting. Using one sample data set with the four AR classes (the selected C-class ARs are marked in Table~\ref{samples}), we tune these arguments using the \verb|tune| function in the R package \verb|e1071| \citep{e1071}, and find that setting \verb|ntree|~$=1000$ and \verb|mtry|~$=6$ usually produces results with slightly better accuracy than those computed with the default argument values (\verb|ntree|~$=500$ and \verb|mtry|~$=\sqrt{F}=3$ in our case). In general, the choice of the \verb|mtry| argument has little impact on the accuracy of RF predictions \citep[e.g.,][]{liaw02}. For demonstration purpose, a sample representative RF tree, created using the R package \verb|reprtree| \citep{reprtree}, is portrayed in Figure~\ref{tree} in appendix.

\begin{figure}[!t]
\epsscale{1.18}
\plotone{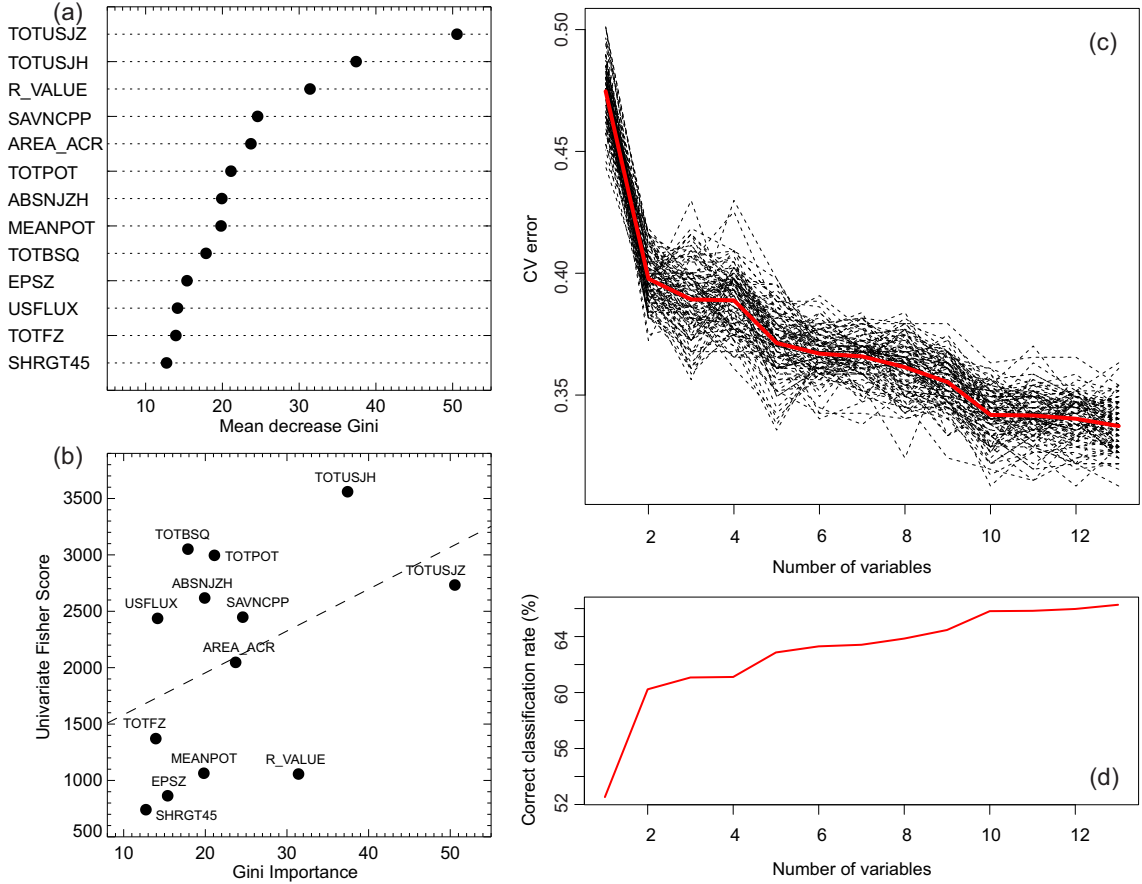}
\caption{Magnetic parameter importance. (a) The Gini importance of the 13 SDO/HMI magnetic parameters for predicting flaring AR classes using the RF method. (b) Univariate Fisher score calculated by \citet{bobra15} vs. Gini importance from RF. The Pearson correlation coefficient is 0.42. The dashed line is a linear fit to the data. (c) 10-fold CV error vs. number of parameters used, showing the result from each of 100 iterations (black dotted line) and average of the 100 iterations (red thick line). (d) Averaged correct classification rate vs. number of parameters used, based on 100 iterations of the 10-fold CV. \label{f2} }
\end{figure}

\subsection{Parameter Importance} \label{features}
As suggested by \citet{bobra15}, the 13 SDO/HMI magnetic parameters listed in Table~\ref{sample_overview} are most useful for discriminating flaring from non-flaring ARs. A natural question is: what is the relative importance of these parameters in classifying flaring ARs into B-, M-, C-, and X-class? An advantage of the RF method is that it can rank the importance of predictive variables by measuring their impacts on accuracy or Gini impurity when splitting nodes \citep{breiman01}. A larger value of Gini importance means that this particular variable plays a greater role in partitioning data into the defined classes. Using the same data set as above, the \verb|randomForest| function generates the Gini importance (mean decrease Gini) of the magnetic parameters as illustrated in Figure~\ref{f2}(a) and also listed in Table~\ref{sample_overview}. The result shows that the order of parameter importance is generally consistent with the correlations of parameters with the AR class (Table~\ref{spearman}), with the total unsigned vertical current (TOTUSJZ), the total unsigned current helicity (TOTUSJH), and the total unsigned flux around high-gradient AR polarity inversion lines (R\_VALUE) being among the most important parameters. We caution that due to the randomness associated with RF (random trees and randomly sampled predictors at each node), different iterations may produce different results, but the order of the most important predictors remains relatively stable. In any case, the parameter rank determined by Gini importance in RF only roughly agrees with that determined by the univariate Fisher score \citep{bobra15}. Figure~\ref{f2}(b) presents a scatter plot of these two quantities for the 13 parameters, showing a weak to moderate Pearson correlation with a coefficient of $\sim$0.42. One parameter of interest is R\_VALUE, which has a relatively low univariate score but a quite high Gini importance. The R\_VALUE parameter was shown to be effective in forecasting major flares \citep{schrijver07,welsch09}. Our result supports this view and connotes that it could also help classifying flaring ARs into different classes. Lastly, we reiterate that the Gini importance derived here is for the purpose of classifying multi-class ARs, different from that of the univariate score of \citet{bobra15} that is to distinguish between flaring and non-flaring ARs.

Based on the variable importance, the function \verb|rfcv| in the \verb|randomForest| package can perform automatic variable selections, by sequentially removing the least important variable(s) and performing the nested cross-validated model prediction. We show in Figure~\ref{f2}(c) and Figure~\ref{f2}(d) the error of 10-fold CV (repeated 100 times) and the averaged correct classification rate, respectively, against the number of SDO/HMI parameters used at each step. It suggests that the optimal model, with an error rate of \sm0.34 and an equivalent overall correct classification rate of \sm66\%, can be achieved for this data set when all 13 variables are used. Similar results are obtained when using other multi-class data sets we constructed. Thus, no further feature selection is carried out, and below we use all the 13 SDO/HMI parameters as input variables of the RF model.

\subsection{Flare Prediction using 13 SDO/HMI Parameters and RF} \label{experiments}
In this subsection, we describe the flare prediction results using the RF algorithm, and compare with previous studies of flare forecasting for a 24 hrs time interval. Using different kinds of flaring AR data sets we prepared, we carry out two experiments, the main one\textsuperscript{\ref{note2}} for multi-class classification (i.e., B-, C-, M-, and X-class ARs) and an additional one for binary classification (i.e., B/C- and M/X-class ARs). The latter is analogous to forecasting flares larger than a certain class (here M1.0 class), which is the approach of most previous studies. For each experiment, as we described earlier we obtain the final results based on the average of 100 times 10-fold CVs on each of the 100 AR samples. The TSS score is used as the standard measure for results comparison, while other metrics are also considered.

\begin{table}[t]
\begin{center}
\caption{RF Multi-class Flare Prediction Results (within 24 hrs) using 13 SDO/HMI Parameters and Comparison to Other Studies\label{result_1}}
\setlength{\tabcolsep}{2.5pt}
\footnotesize
\begin{tabular}{l|cccc}
\hline\noalign{\smallskip}
Prediction$\downarrow$ Observation$\rightarrow$ & B Class (n=128) & C Class (n=142) & M Class (n=142) & X Class (n=23)\\
\hline\noalign{\smallskip}
B Class & 103.980$\pm$3.624  &    33.413$\pm$4.287  &    10.467$\pm$1.770 &     0.000$\pm$0.000 \\
C Class & 23.187$\pm$3.569  &    74.700$\pm$5.685   &   33.984$\pm$3.724  &   0.884$\pm$0.492 \\
M Class & 0.833$\pm$0.839  &    33.866$\pm$4.259  &    95.237$\pm$3.703  &    15.296$\pm$0.939 \\
X Class & 0.000$\pm$0.000 &  0.021$\pm$0.144  &    2.312$\pm$1.294   &   6.819$\pm$0.804 \\ \hline
Recall &  &   &   &   \\
This work & 0.812$\pm$0.039 & 0.526$\pm$0.050 & 0.671$\pm$0.037 & 0.297$\pm$0.039 \\
\citet{yuan10} & 0.714 & 0.138 & 0.221 & 0.206 \\ 
\citet{bloomfield12} & N/A & 0.737 & 0.693 & 0.859 \\ \hline
Precision &  &   &   &   \\
This work & 0.703$\pm$0.037 & 0.563$\pm$0.054 & 0.656$\pm$0.036 & 0.745$\pm$0.152 \\
\citet{yuan10} & 0.763 & 0.529 & 0.357 & 0.438 \\
\citet{bloomfield12} & N/A & 0.330 & 0.136 & 0.029 \\ \hline
Accuracy &  &   &   &   \\
This work & 0.844$\pm$0.017 & 0.712$\pm$0.026 & 0.778$\pm$0.019 & 0.957$\pm$0.005 \\
\citet{yuan10} & 0.861 & 0.722 & 0.652 & 0.843 \\
\citet{bloomfield12} & N/A & 0.711 & 0.829 & 0.881 \\ \hline
TSS &  &   &   &   \\
This work & 0.669$\pm$0.039 & 0.328$\pm$0.050 & 0.500$\pm$0.037 & 0.291$\pm$0.039 \\
\citet{yuan10} & 0.630 & 0.090 & 0.054 & 0.160 \\
\citet{bloomfield12} & N/A & 0.443 & 0.526 & 0.740 \\ \hline
\noalign{\smallskip}\hline
\end{tabular}
\end{center}
\tablecomments{For \citet{yuan10}, we use the contingency tables they provide in Figures 3--6. For \citet{bloomfield12}, we use their Table~4 and also retrieve the contingency table from the machine-readable data they provide online.}
\end{table}

\begin{table}[t]
\begin{center}
\caption{RF Binary-class Flare Prediction Results (within 24 hrs) using 13 SDO/HMI Parameters and Comparison to Other Studies\label{result_2}}
\footnotesize
\begin{tabular}{l|cc}
\hline\noalign{\smallskip}
Prediction$\downarrow$ Observation$\rightarrow$ & B/C Class (n=165) & M/X Class (n=165)\\
\hline\noalign{\smallskip}
B/C Class & 129.536$\pm$4.025  &    41.722$\pm$3.370  \\
M/X Class & 35.464$\pm$4.025 &  123.278$\pm$3.370 \\ \hline
Recall &  &   \\
This work & 0.785$\pm$0.036 & 0.747$\pm$0.030 \\
\citet{bloomfield12} & N/A & 0.704  \\
\citet{ahmed13} & N/A & 0.523  \\
\citet{bobra15} & N/A & 0.832$\pm$0.042 \\
\citet{nishizuka17} & N/A & 0.716 \\ \hline
Precision &  &   \\
This work & 0.756$\pm$0.033 & 0.777$\pm$0.033 \\
\citet{bloomfield12} & N/A & 0.146 \\
\citet{ahmed13} & N/A & 0.740  \\
\citet{bobra15} & N/A & 0.417$\pm$0.037 \\
\citet{nishizuka17} & N/A & 0.969 \\ \hline
Accuracy &  &   \\
This work & 0.766$\pm$0.023 & 0.766$\pm$0.021 \\
\citet{bloomfield12} & N/A & 0.830 \\ 
\citet{ahmed13} & N/A & 0.963  \\
\citet{bobra15} & N/A & 0.924$\pm$0.007 \\ 
\citet{nishizuka17} & N/A & 0.990 \\ \hline
TSS &  &    \\
This work & 0.532$\pm$0.036 & 0.532$\pm$0.030 \\
\citet{bloomfield12} & N/A & 0.539 \\
\citet{ahmed13} & N/A & 0.512  \\
\citet{bobra15} & N/A & 0.761$\pm$0.039 \\
\citet{nishizuka17} & N/A & 0.71$\pm$0.002 \\ \hline
\noalign{\smallskip}\hline
\end{tabular}
\end{center}
\tablecomments{The comparison is made between our results and previous works on forecasting flares larger than M1.0 class. For \citet{bloomfield12}, we use their Table~4 and also retrieve the contingency table from the machine-readable data they provide online. For \citet{ahmed13} and \citet{bobra15}, we use the results provided by \citet{bobra15} in their Table~2. For \citet{nishizuka17}, we use the contingency table they provide in Table~3 and only the entries for the ERT method.}
\end{table}

First, we apply RF to predict flaring ARs with four different classes. Detailed results including a confusion matrix and model performance metrics are presented in Table~\ref{result_1}. It can be seen that the RF method works reasonably well in terms of recall, precision, and accuracy, except for a low recall of X-class (\sm0.30). A closer look reveals that the RF has difficulties in distinguishing X- from M-class AR samples. This is presumably due to the sparsity of the current X-class samples available for training, as the solar cycle 24 contemporaneous with SDO/HMI measurements is unusually quiet compared to previous cycles. Nevertheless, our TSS scores (about 0.67 for B class, 0.33 for C class, 0.50 for M class, and 0.29 for X class) outperform \citet{yuan10} in every AR class; in particular, we obtain a \sm10 times higher TSS score for the M class. Compared to \citet{bloomfield12} which used solar cycles 21 and 22 observations for training, our TSS scores of the C- and M-classes are roughly similar while that of the X-class is \sm2.5 times lower. However, it can be noted that the precision of X-class of \citet{bloomfield12} is very low (due to a large number of FP counts), about 26 times lower than ours, while at the mean time, their recall of the X-class is only about 3 times better. We recognize that recall is a very important metric especially for the most energetic X-class ARs/flares, as a miss is usually deemed worse than a false alarm. It could be expected that provided with more X-class training samples, our method may yield improved recall performance while maintaining a high precision prediction for X-class ARs.

Second, we make binary predictions and show the confusion matrix and performance metrics in Table~\ref{result_2}. We also compare our results with previous works on forecasting flares larger than M1.0 class. By looking at the TSS score, it is apparent that the performance of our method coupling SDO/HMI magnetic parameters with RF (TSS~$\approx 0.53$) is very similar to \citet{bloomfield12} and \citet{ahmed13}, while being \sm28\% lower than \citet{bobra15} and \citet{nishizuka17}. Nonetheless, we obtain quite high scores in both recall and precision. We note that \citet{nishizuka17} used the ERT classifier similar to RF, and also considered, besides magnetic field parameters, UV brightenings and previous flare activity to achieve high metric scores.

It is worth noting that all our training AR samples are flare productive, with the AR classes defined in the same way as \citet{yuan10} (see Section~\ref{data}). Many previous studies \citep[e.g.,][]{ahmed13,bobra15,nishizuka17} include non-flaring ARs as negative samples. This difference in building data sets with multiple classes may affect the performance comparisons. For example, the database for our binary classification is constructed from multi-class AR samples. As a result, the RF classifier actually works to distinguish M/X-class ARs from B/C-class ARs (\sm80\% are C-class ARs in our data sets). Thus intuitively, including non-flaring AR samples might facilitate/improve our prediction. The ideal situation would be to conduct performance comparisons between different flare forecasting methods from the same database \citep{barnes08}. 

\section{SUMMARY AND DISCUSSIONS} \label{summary}
Although not based on physical models of flares, solar flare prediction taking advantage of modern machine learning techniques has drawn significant attention in recent years. What distinguishes the present work from previous studies is the use of prediction parameters from the state-of-the-art SDO/HMI instrument and the advanced RF algorithm, with the main goal of multi-class flare forecasting. Based on flare events that occur from 2010 May to 2016 December, we build a database containing sample ARs that belong to four classes (B, C, M, and X), stipulated according to the maximum GOES class of flare(s) that an AR ever produces. We then apply the RF classifier to make flare predictions in 24 hours, evaluate the model performance using the 10-fold CV technique, and characterize our results and make comparisons with previous studies using performance metrics. The main results are summarized as follows.

\begin{enumerate}  

\item From the evaluation of variable importance by the RF algorithm, the 13 SDO/HMI magnetic parameters used by \citet{bobra15} to discriminate flaring and non-flaring ARs are also helpful in distinguishing flaring ARs into four different classes, yet the orders of parameter importance determined from these two approaches only roughly agree, with a weak to moderate Pearson correlation of \sm0.42. The three most important parameters for classifying ARs are TOTUSJZ, TOTUSJH, and R\_VALUE. We surmise that the ranking of one parameter versus another is not really as significant as using all 13 parameters in combination.

\item In classifying flaring ARs into multiple classes, we achieve a TSS score of about 0.70, 0.33, 0.50, and 0.29 for B-, C-, M-, and X-class ARs, respectively, which clearly outperform \citet{yuan10} that used the same way for defining AR samples. Our TSS scores of C and M classes are roughly comparable to those of \citet{bloomfield12}, but that of the X class is \sm2.5 times lower, most probably due to the lack of X-class training samples in the solar cycle 24. In forecasting flares larger than M1.0 class, our method yields a TSS score of \sm0.53, comparable to \citet{bloomfield12} and \cite{ahmed13}, while being \sm28\% lower than \citet{bobra15} and \citet{nishizuka17}. In general, we obtain fairly good scores in recall and precision at the same time.

\end{enumerate}

It should be noted that (1) these result comparisons could be affected by the fact that the sample classes in the present study are defined using different criteria from many previous studies, and we consider flaring ARs only without including non-flaring AR regions. (2) To better train the random forest model for the main objective of multi-class classification, we randomly select C-class samples from a larger pool to construct training data with more balanced AR classes. Indeed, if we use the originally obtained 845 AR samples without undersampling the C class, the resulted TSS scores after 100 times of CVs are 0.411$\pm$0.023, 0.366$\pm$0.011, 0.341$\pm$0.021, and 0.268$\pm$0.025 for the B, C, M, and X class, respectively, much inferior to those in Table~\ref{result_1} except for the C class. Then the question is whether the model developed with the undersampled C class data can perform equally well for test data with original class ratios. To answer this, we revise the main experiment in Section~\ref{experiments} as follows. After randomly selecting 142 unique C-class ARs from the total of 552 C-class samples to form a complete data set with samples in other classes, we keep the leftover 410 C-class samples; when carrying out the 10-fold CVs, the data for validation also includes one fold from the 410 C-class data. In doing so, we build the model with undersampled C-class data (same as the previous main experiment) but validate it using data that reflects the original class ratios. The TSS scores obtained this way are 0.620$\pm$0.039, 0.333$\pm$0.028, 0.463$\pm$0.036, and 0.294$\pm$0.039 for the B, C, M, and X class, respectively, very similar to those in Table~\ref{result_1} (with difference only up to \sm7\%). All these show that for this study, using data with more balanced classes can help build the optimal model, which can also perform well in an operational setting.

Based on all our experiments, we conclude that using SDO/HMI magnetic parameters and the RF algorithm is a valid method for flare forecasting. Importantly, incorporating other features of flaring ARs, such as previous flare activity, could be critical in improving the prediction performance \citep{nishizuka17}. Extended studies should also be made on multi-class flare forecasting using ERT, as it could be more computationally efficient and robust compared to RF. Related to multi-class classification, it will also be interesting to find out the most useful parameter for predicting a certain flare class. Another possible future work is to calculate the flare index \citep{abramenko05} for each AR, which can reflect the overall flare productivity by taking into account the numbers of flares of different GOES classes within a certain time period; then the RF/ERT classifier can be run in the regression mode to predict the flare index quantitatively. More generally, as solar data from various instruments are growing rapidly, including such as those from SDO and the recently digitized historical ground-based H-alpha images covering many solar cycles \citep{liu10b}, the RF/ERT algorithms may be helpful in solving other multi-class problems in solar physics.

\acknowledgments
We thank the team of SDO/HMI for producing vector magnetic field data products, and the anonymous referee for valuable comments that helped us improve this work. The ``X-ray Flare'' data set was prepared by and made available through NOAA NCEI. R software and the packages used are available from CRAN (\url{http://CRAN.R-project.org/}). C.L., N.D., and H.W. acknowledge NASA grant NNX16AD67G for support of efforts related to solar big data processing.

\vspace{5mm}
\facility{SDO (HMI)}

\clearpage
\appendix

\startlongtable
\begin{deluxetable*}{ccc|ccc|ccc}
\tabletypesize{\scriptsize}
\tablecolumns{9}
\tablewidth{0pt}
\tablecaption{845 Samples of Flaring ARs \label{samples}} 
\tablehead{
\colhead{Start Time} & \colhead{NOAA} & \colhead{GOES} & \colhead{Start Time} & \colhead{NOAA} & \colhead{GOES} & \colhead{Start Time} & \colhead{NOAA} & \colhead{GOES}\\
\colhead{YY/MM/DD UT} & \colhead{AR} & \colhead{Class} & \colhead{YY/MM/DD UT} & \colhead{AR} & \colhead{Class} & \colhead{YY/MM/DD UT} & \colhead{AR} & \colhead{Class}} 
\startdata
11/02/15 01:44 & 11158 & X2.2 & 11/03/09 23:13 & 11166 & X1.5 & 11/09/06 22:12 & 11283 & X2.1 \\
11/09/07 22:32 & 11283 & X1.8 & 11/09/24 09:21 & 11302 & X1.9 & 12/03/05 02:30 & 11429 & X1.1 \\
12/03/07 00:02 & 11429 & X5.4 & 12/07/12 15:37 & 11520 & X1.4 & 13/05/15 01:25 & 11748 & X1.2 \\
13/10/28 01:41 & 11875 & X1.0 & 13/11/05 22:07 & 11890 & X3.3 & 13/11/08 04:20 & 11890 & X1.1 \\
13/11/10 05:08 & 11890 & X1.1 & 13/11/19 10:14 & 11893 & X1.0 & 14/01/07 18:04 & 11944 & X1.2 \\
14/03/29 17:35 & 12017 & X1.0 & 14/10/22 14:02 & 12192 & X1.6 & 14/10/25 16:55 & 12192 & X1.0 \\
14/10/26 10:04 & 12192 & X2.0 & 14/10/27 14:12 & 12192 & X2.0 & 14/10/24 21:07 & 12192 & X3.1 \\
14/11/07 16:53 & 12205 & X1.6 & 15/03/11 16:11 & 12297 & X2.2 & 10/05/05 17:13 & 11069 & M1.2 \\
10/06/12 00:30 & 11081 & M2.0 & 10/08/07 17:55 & 11093 & M1.0 & 10/10/16 19:07 & 11112 & M2.9 \\
10/11/06 15:27 & 11121 & M5.4 & 11/03/07 09:14 & 11164 & M1.8 & 11/03/07 19:43 & 11165 & M3.7 \\
11/03/14 19:30 & 11169 & M4.2 & 11/03/24 12:01 & 11176 & M1.0 & 11/06/07 06:16 & 11226 & M2.5 \\
11/07/27 15:48 & 11260 & M1.1 & 11/08/03 03:08 & 11261 & M1.1 & 11/08/04 03:41 & 11261 & M9.3 \\
11/09/23 01:47 & 11295 & M1.6 & 11/09/30 18:55 & 11305 & M1.0 & 11/10/01 08:56 & 11305 & M1.2 \\
11/10/02 00:37 & 11305 & M3.9 & 11/12/25 18:11 & 11380 & M4.0 & 11/12/26 02:13 & 11380 & M1.5 \\
12/02/06 19:31 & 11410 & M1.0 & 12/03/14 15:08 & 11432 & M2.8 & 12/03/15 07:23 & 11432 & M1.8 \\
12/03/17 20:32 & 11434 & M1.4 & 12/04/27 08:15 & 11466 & M1.0 & 12/05/07 14:03 & 11471 & M1.9 \\
12/05/08 13:02 & 11476 & M1.4 & 12/05/09 14:02 & 11476 & M1.8 & 12/05/10 04:11 & 11476 & M5.7 \\
12/05/06 17:41 & 11476 & M1.3 & 12/06/06 19:54 & 11494 & M2.1 & 12/06/14 12:52 & 11504 & M1.9 \\
12/07/04 16:33 & 11513 & M1.8 & 12/07/06 13:26 & 11513 & M1.2 & 12/07/02 19:59 & 11515 & M3.8 \\
12/07/04 14:35 & 11515 & M1.3 & 12/07/05 20:09 & 11515 & M1.6 & 12/07/06 18:48 & 11515 & M1.3 \\
12/07/07 10:57 & 11515 & M2.6 & 12/07/29 06:15 & 11532 & M2.3 & 12/07/30 15:39 & 11532 & M1.1 \\
12/07/28 20:44 & 11532 & M6.1 & 12/08/18 00:24 & 11543 & M5.5 & 12/09/09 21:50 & 11564 & M1.2 \\
12/11/20 19:21 & 11618 & M1.6 & 12/11/21 06:45 & 11618 & M1.4 & 12/11/27 21:05 & 11620 & M1.0 \\
12/11/28 21:20 & 11620 & M2.2 & 13/01/11 14:51 & 11654 & M1.0 & 13/02/17 15:45 & 11675 & M1.9 \\
13/03/05 07:47 & 11686 & M1.2 & 13/03/15 05:46 & 11692 & M1.1 & 13/04/11 06:55 & 11719 & M6.5 \\
13/05/02 04:58 & 11731 & M1.1 & 13/05/03 16:39 & 11731 & M1.3 & 13/05/05 17:42 & 11734 & M1.4 \\
13/05/31 19:52 & 11760 & M1.0 & 13/06/05 08:14 & 11762 & M1.3 & 13/06/23 20:48 & 11778 & M2.9 \\
13/08/12 10:21 & 11817 & M1.5 & 13/08/17 18:49 & 11818 & M1.4 & 13/10/24 10:30 & 11877 & M3.5 \\
13/10/28 04:32 & 11877 & M5.1 & 13/10/26 19:49 & 11882 & M1.0 & 13/10/28 15:07 & 11882 & M4.4 \\
13/10/25 20:54 & 11882 & M1.9 & 13/11/01 19:46 & 11884 & M6.3 & 13/11/02 22:13 & 11884 & M1.6 \\
13/11/03 05:16 & 11884 & M4.9 & 13/11/08 09:22 & 11891 & M2.3 & 13/11/15 02:20 & 11899 & M1.0 \\
13/11/23 12:49 & 11899 & M1.0 & 13/12/07 07:17 & 11909 & M1.2 & 13/12/22 21:23 & 11928 & M1.6 \\
13/12/29 07:49 & 11936 & M3.1 & 13/12/31 21:45 & 11936 & M6.4 & 14/01/01 18:40 & 11936 & M9.9 \\
14/01/07 03:49 & 11946 & M1.0 & 14/02/01 07:14 & 11967 & M3.0 & 14/02/02 09:24 & 11967 & M4.4 \\
14/02/04 15:25 & 11967 & M1.5 & 14/02/07 04:47 & 11967 & M2.0 & 14/02/02 06:24 & 11968 & M2.6 \\
14/02/04 01:16 & 11968 & M3.8 & 14/02/07 10:25 & 11968 & M1.9 & 14/02/11 16:34 & 11974 & M1.8 \\
14/02/12 06:54 & 11974 & M2.3 & 14/02/13 15:45 & 11974 & M1.4 & 14/02/14 16:33 & 11974 & M1.0 \\
14/02/16 09:20 & 11977 & M1.2 & 14/02/20 07:26 & 11982 & M3.0 & 14/02/28 00:44 & 11991 & M1.1 \\
14/03/05 02:06 & 11991 & M1.0 & 14/03/10 22:45 & 11996 & M1.4 & 14/03/11 03:44 & 11996 & M3.5 \\
14/03/09 20:13 & 12002 & M1.0 & 14/03/10 15:21 & 12002 & M1.7 & 14/04/16 19:54 & 12035 & M1.0 \\
14/06/03 03:58 & 12077 & M1.4 & 14/06/12 09:23 & 12085 & M1.8 & 14/06/12 19:56 & 12089 & M1.1 \\
14/07/08 16:06 & 12113 & M6.5 & 14/10/09 06:48 & 12182 & M1.2 & 14/12/01 06:26 & 12222 & M1.8 \\
14/12/04 08:00 & 12222 & M1.3 & 14/12/17 18:54 & 12241 & M1.4 & 14/12/18 21:41 & 12241 & M6.9 \\
14/12/27 02:03 & 12249 & M2.2 & 15/01/03 09:40 & 12253 & M1.1 & 15/01/04 15:18 & 12253 & M1.3 \\
15/01/28 04:21 & 12268 & M1.4 & 15/01/29 11:32 & 12268 & M2.1 & 15/01/30 05:29 & 12277 & M1.7 \\
15/02/04 02:08 & 12277 & M1.2 & 15/02/09 22:19 & 12280 & M2.4 & 15/04/08 14:37 & 12320 & M1.4 \\
15/05/05 17:12 & 12335 & M2.6 & 15/06/21 18:10 & 12367 & M1.1 & 15/06/20 06:28 & 12371 & M1.0 \\
15/06/21 01:02 & 12371 & M2.0 & 15/06/22 17:39 & 12371 & M6.6 & 15/06/25 08:02 & 12371 & M7.9 \\
15/07/03 12:47 & 12378 & M1.5 & 15/07/06 20:32 & 12381 & M1.7 & 15/08/21 19:10 & 12403 & M1.1 \\
15/08/22 21:19 & 12403 & M3.5 & 15/08/24 07:26 & 12403 & M5.6 & 15/08/27 04:48 & 12403 & M2.9 \\
15/08/28 13:04 & 12403 & M2.2 & 15/09/17 09:34 & 12415 & M1.1 & 15/09/20 17:32 & 12415 & M2.1 \\
15/10/01 13:03 & 12422 & M4.5 & 15/10/16 06:11 & 12434 & M1.1 & 15/10/15 23:27 & 12434 & M1.1 \\
15/10/31 17:48 & 12443 & M1.0 & 15/11/04 13:31 & 12443 & M3.7 & 15/11/09 12:49 & 12449 & M3.9 \\
15/12/23 00:23 & 12473 & M4.7 & 16/02/13 15:16 & 12497 & M1.8 & 16/02/14 19:18 & 12497 & M1.0 \\
16/02/15 10:41 & 12497 & M1.2 & 16/04/18 00:14 & 12529 & M6.7 & 16/11/29 23:29 & 12615 & M1.2 \\
10/07/13 10:43 & 11087 & C2.6 & 10/08/01 07:55 & 11092 & C3.2 & 10/09/06 14:54 & 11105 & C2.5$^{*}$ \\
10/10/27 16:59 & 11117 & C1.2 & 10/10/31 03:13 & 11117 & C1.8 & 10/10/26 08:09 & 11119 & C1.0 \\
10/11/15 07:28 & 11124 & C2.3 & 10/12/14 15:03 & 11133 & C2.3$^{*}$ & 11/01/03 23:26 & 11142 & C1.1 \\
11/03/27 23:18 & 11181 & C1.0 & 11/04/18 18:55 & 11193 & C1.5 & 11/04/29 20:40 & 11199 & C1.8$^{*}$ \\
11/05/15 14:11 & 11208 & C1.3 & 11/05/09 17:31 & 11210 & C1.5 & 11/06/15 14:19 & 11234 & C3.2 \\
11/06/17 23:37 & 11234 & C3.9 & 11/06/19 16:03 & 11237 & C1.5$^{*}$ & 11/07/07 02:31 & 11243 & C1.0 \\
11/07/03 10:54 & 11244 & C2.1 & 11/07/08 14:56 & 11247 & C2.3$^{*}$ & 11/07/11 10:47 & 11247 & C2.6 \\
11/07/12 14:44 & 11247 & C1.9 & 11/07/18 10:19 & 11254 & C1.0 & 11/07/30 12:11 & 11265 & C1.3 \\
11/07/31 19:01 & 11265 & C1.7 & 11/08/09 13:29 & 11266 & C2.2 & 11/08/05 05:56 & 11267 & C1.3 \\
11/08/06 11:41 & 11267 & C1.3 & 11/08/07 08:24 & 11267 & C1.6 & 11/08/18 14:59 & 11271 & C1.1 \\
11/08/20 22:54 & 11271 & C2.9$^{*}$ & 11/08/24 16:31 & 11271 & C1.1 & 11/08/26 03:29 & 11271 & C1.0$^{*}$ \\
11/08/17 16:16 & 11272 & C2.6 & 11/08/18 06:40 & 11272 & C1.2 & 11/08/21 18:15 & 11272 & C1.5 \\
11/08/30 03:06 & 11274 & C1.9 & 11/09/03 03:19 & 11280 & C1.0 & 11/08/30 22:02 & 11281 & C5.5 \\
11/09/02 15:09 & 11281 & C1.8$^{*}$ & 11/09/03 13:57 & 11281 & C1.0$^{*}$ & 11/09/08 18:26 & 11289 & C2.5 \\
11/09/16 22:09 & 11290 & C1.4$^{*}$ & 11/09/14 20:42 & 11297 & C9.2 & 11/09/15 21:09 & 11297 & C2.6$^{*}$ \\
11/09/16 08:42 & 11297 & C2.2$^{*}$ & 11/09/20 06:44 & 11301 & C2.0$^{*}$ & 11/09/21 11:17 & 11301 & C3.9 \\
11/10/07 01:14 & 11313 & C1.2$^{*}$ & 11/10/10 14:30 & 11313 & C4.5$^{*}$ & 11/10/11 23:19 & 11316 & C1.1 \\
11/10/12 09:46 & 11316 & C1.7 & 11/10/14 15:09 & 11316 & C1.1 & 11/10/15 18:13 & 11316 & C1.5 \\
11/10/16 13:51 & 11317 & C1.4 & 11/10/22 15:14 & 11324 & C4.1 & 11/10/28 11:48 & 11324 & C1.7 \\
11/10/30 09:24 & 11330 & C2.4 & 11/11/16 23:40 & 11346 & C5.0 & 11/11/17 07:16 & 11346 & C6.0 \\
11/11/18 16:38 & 11346 & C1.1$^{*}$ & 11/11/17 13:53 & 11352 & C3.1 & 11/11/18 17:07 & 11354 & C2.7$^{*}$ \\
11/11/19 11:19 & 11354 & C1.4 & 11/11/20 16:35 & 11354 & C6.1 & 11/11/22 00:33 & 11354 & C1.2 \\
11/11/26 17:12 & 11358 & C1.0 & 11/11/27 11:59 & 11358 & C1.1$^{*}$ & 11/11/25 21:49 & 11359 & C2.4 \\
11/11/28 18:22 & 11361 & C3.2$^{*}$ & 11/11/29 03:25 & 11361 & C2.1 & 11/12/01 20:27 & 11361 & C1.1 \\
11/12/05 03:09 & 11361 & C1.9 & 11/11/29 08:54 & 11362 & C2.5$^{*}$ & 11/12/05 06:35 & 11362 & C1.4 \\
11/12/03 05:48 & 11363 & C1.2 & 11/12/05 23:20 & 11363 & C6.9 & 11/12/07 15:34 & 11364 & C1.3 \\
11/12/09 13:05 & 11374 & C3.1$^{*}$ & 11/12/10 23:00 & 11374 & C1.0 & 11/12/12 08:30 & 11374 & C1.2 \\
11/12/17 22:47 & 11376 & C3.2 & 11/12/18 04:59 & 11376 & C1.8 & 11/12/20 22:38 & 11376 & C6.2 \\
11/12/21 20:08 & 11376 & C1.5$^{*}$ & 11/12/22 01:56 & 11381 & C5.4$^{*}$ & 11/12/27 04:11 & 11386 & C8.9 \\
11/12/28 20:18 & 11386 & C4.0 & 11/12/31 17:17 & 11386 & C1.2 & 12/01/10 22:34 & 11391 & C1.7 \\
12/01/13 06:01 & 11391 & C2.2 & 12/01/08 06:08 & 11393 & C1.9 & 12/01/09 10:31 & 11393 & C1.1$^{*}$ \\
12/01/10 01:49 & 11393 & C1.0 & 12/01/09 20:01 & 11395 & C2.6 & 12/01/11 11:04 & 11395 & C1.6 \\
12/01/12 00:49 & 11395 & C1.5$^{*}$ & 12/01/14 03:19 & 11396 & C2.1 & 12/01/19 12:41 & 11396 & C3.2$^{*}$ \\
12/02/08 21:48 & 11415 & C2.9 & 12/02/19 08:41 & 11422 & C1.0$^{*}$ & 12/03/01 15:10 & 11423 & C3.4 \\
12/03/03 03:01 & 11427 & C1.3 & 12/03/04 17:24 & 11427 & C3.2 & 12/03/08 02:49 & 11428 & C7.2$^{*}$ \\
12/03/10 06:53 & 11428 & C1.9 & 12/03/21 13:54 & 11440 & C1.2 & 12/04/04 16:17 & 11450 & C1.2 \\
12/04/05 20:49 & 11450 & C1.5 & 12/04/18 16:54 & 11459 & C5.2$^{*}$ & 12/04/21 20:10 & 11459 & C1.8$^{*}$ \\
12/04/22 21:05 & 11459 & C1.7 & 12/04/19 00:00 & 11460 & C1.4 & 12/04/21 01:27 & 11460 & C2.4$^{*}$ \\
12/04/23 17:38 & 11461 & C2.0 & 12/04/18 14:51 & 11463 & C5.9 & 12/04/19 02:54 & 11463 & C2.9 \\
12/04/20 22:25 & 11465 & C1.4 & 12/04/22 21:42 & 11465 & C2.4 & 12/04/27 10:53 & 11465 & C2.4 \\
12/04/27 13:18 & 11467 & C2.0 & 12/04/28 21:48 & 11467 & C1.1 & 12/04/27 01:59 & 11469 & C1.0$^{*}$ \\
12/04/28 08:54 & 11469 & C1.7 & 12/04/29 14:27 & 11469 & C1.1 & 12/05/02 20:38 & 11469 & C1.9$^{*}$ \\
12/05/03 16:39 & 11469 & C2.3 & 12/05/04 04:48 & 11469 & C1.3 & 12/05/19 12:04 & 11479 & C1.0 \\
12/05/14 19:22 & 11483 & C1.8 & 12/05/16 00:14 & 11484 & C1.8$^{*}$ & 12/05/23 00:15 & 11484 & C1.2 \\
12/05/24 19:57 & 11488 & C3.9 & 12/05/27 04:42 & 11492 & C3.1 & 12/06/12 05:47 & 11506 & C1.0 \\
12/06/11 18:59 & 11507 & C1.5$^{*}$ & 12/06/12 16:01 & 11507 & C1.4$^{*}$ & 12/06/25 20:55 & 11512 & C1.4 \\
12/06/27 07:57 & 11512 & C3.2 & 12/06/28 04:45 & 11512 & C4.2$^{*}$ & 12/06/29 18:35 & 11512 & C1.1 \\
12/07/27 03:58 & 11528 & C5.0 & 12/08/01 12:47 & 11535 & C2.3 & 12/08/03 21:20 & 11535 & C3.0 \\
12/08/04 06:03 & 11535 & C1.3 & 12/08/09 05:27 & 11538 & C3.2 & 12/08/08 16:10 & 11542 & C1.9 \\
12/08/09 22:42 & 11542 & C1.4 & 12/08/10 18:38 & 11542 & C1.4 & 12/08/11 16:28 & 11542 & C2.0$^{*}$ \\
12/08/14 22:23 & 11542 & C1.6 & 12/09/01 19:39 & 11553 & C1.9 & 12/09/02 13:37 & 11553 & C1.1$^{*}$ \\
12/08/25 02:24 & 11554 & C1.7 & 12/08/30 04:48 & 11554 & C1.5 & 12/08/31 19:45 & 11560 & C8.4 \\
12/09/02 18:00 & 11560 & C5.5$^{*}$ & 12/09/03 16:30 & 11560 & C1.3$^{*}$ & 12/09/05 16:14 & 11560 & C1.1 \\
12/09/06 03:11 & 11560 & C1.8$^{*}$ & 12/09/07 17:05 & 11562 & C1.3 & 12/09/11 08:25 & 11567 & C1.1 \\
12/09/09 14:51 & 11568 & C1.7$^{*}$ & 12/09/11 01:00 & 11569 & C3.5 & 12/09/25 06:34 & 11573 & C1.0 \\
12/09/20 11:27 & 11574 & C1.0 & 12/09/25 04:24 & 11577 & C3.6 & 12/10/10 21:40 & 11585 & C1.5 \\
12/10/17 00:44 & 11589 & C1.1 & 12/10/19 05:10 & 11589 & C1.9 & 12/10/23 07:40 & 11593 & C3.0 \\
12/10/19 18:42 & 11594 & C1.7$^{*}$ & 12/10/21 02:54 & 11596 & C7.8$^{*}$ & 12/10/26 10:23 & 11596 & C1.5 \\
12/11/14 13:35 & 11611 & C4.3 & 12/11/14 23:33 & 11613 & C1.3 & 12/11/15 20:14 & 11613 & C1.0$^{*}$ \\
12/11/14 13:22 & 11614 & C1.3$^{*}$ & 12/11/29 11:13 & 11623 & C2.0 & 12/12/18 16:37 & 11633 & C1.3 \\
12/12/23 05:43 & 11633 & C1.3$^{*}$ & 13/01/01 08:47 & 11640 & C1.2 & 13/01/12 21:57 & 11652 & C3.1 \\
13/01/13 18:30 & 11652 & C2.3 & 13/01/14 01:15 & 11652 & C6.5 & 13/01/31 04:30 & 11663 & C1.1 \\
13/02/05 08:12 & 11669 & C6.3$^{*}$ & 13/02/06 05:53 & 11669 & C1.0$^{*}$ & 13/02/12 17:47 & 11670 & C1.5$^{*}$ \\
13/03/12 22:42 & 11689 & C3.6 & 13/03/14 05:45 & 11691 & C2.1 & 13/03/19 13:50 & 11695 & C1.5$^{*}$ \\
13/03/13 16:15 & 11696 & C1.4 & 13/03/15 15:09 & 11696 & C1.0 & 13/03/16 08:30 & 11698 & C1.5 \\
13/04/03 18:34 & 11711 & C1.7 & 13/04/07 16:49 & 11713 & C1.6 & 13/04/08 15:42 & 11714 & C1.6 \\
13/04/06 23:31 & 11718 & C1.0 & 13/04/07 15:55 & 11718 & C3.1 & 13/04/09 13:28 & 11718 & C1.5 \\
13/04/10 17:56 & 11718 & C3.4$^{*}$ & 13/04/11 16:52 & 11718 & C1.0 & 13/04/13 19:46 & 11718 & C1.4 \\
13/04/10 12:54 & 11721 & C1.2 & 13/04/11 10:09 & 11721 & C4.2 & 13/04/15 13:58 & 11723 & C1.3 \\
13/04/16 07:30 & 11723 & C1.3 & 13/04/20 22:46 & 11726 & C1.3$^{*}$ & 13/04/21 19:59 & 11726 & C2.7 \\
13/04/23 23:21 & 11726 & C2.5$^{*}$ & 13/04/24 22:39 & 11726 & C1.6 & 13/04/30 20:43 & 11730 & C2.6 \\
13/05/01 07:23 & 11730 & C5.5$^{*}$ & 13/04/29 19:26 & 11733 & C4.0 & 13/05/05 19:57 & 11739 & C8.3 \\
13/05/06 01:58 & 11739 & C2.4 & 13/05/10 16:30 & 11739 & C2.5 & 13/05/11 02:52 & 11744 & C1.3 \\
13/05/17 04:37 & 11744 & C5.0$^{*}$ & 13/05/13 21:58 & 11745 & C8.3 & 13/05/21 10:23 & 11745 & C1.2 \\
13/05/19 21:11 & 11750 & C4.7 & 13/05/20 00:41 & 11750 & C4.0 & 13/05/18 06:32 & 11752 & C1.3 \\
13/05/23 12:26 & 11755 & C1.3 & 13/05/25 20:17 & 11755 & C1.6 & 13/05/21 22:32 & 11756 & C2.9 \\
13/05/22 19:46 & 11756 & C2.4$^{*}$ & 13/05/23 18:41 & 11756 & C3.4 & 13/05/24 16:31 & 11756 & C2.1$^{*}$ \\
13/06/10 14:18 & 11765 & C1.9 & 13/06/19 11:49 & 11773 & C1.5 & 13/06/22 17:31 & 11773 & C1.5 \\
13/06/17 04:29 & 11775 & C1.0 & 13/06/18 06:48 & 11775 & C2.2 & 13/06/19 07:20 & 11775 & C3.5 \\
13/06/20 15:55 & 11775 & C1.4 & 13/06/24 20:35 & 11775 & C3.0$^{*}$ & 13/06/19 16:46 & 11776 & C1.0 \\
13/06/21 18:02 & 11776 & C1.0 & 13/06/30 16:44 & 11780 & C2.0$^{*}$ & 13/06/29 16:01 & 11781 & C1.1 \\
13/07/06 17:31 & 11784 & C1.1 & 13/07/04 05:20 & 11785 & C3.9$^{*}$ & 13/07/05 21:30 & 11785 & C3.5 \\
13/07/06 08:07 & 11785 & C1.0$^{*}$ & 13/07/07 12:05 & 11785 & C1.7 & 13/07/09 13:25 & 11785 & C2.3 \\
13/07/11 09:04 & 11785 & C1.0 & 13/07/13 05:36 & 11791 & C1.4 & 13/07/15 16:55 & 11791 & C1.0$^{*}$ \\
13/07/16 10:10 & 11791 & C1.9$^{*}$ & 13/07/17 01:49 & 11791 & C1.1 & 13/07/20 03:34 & 11793 & C2.1 \\
13/07/21 08:24 & 11800 & C3.1 & 13/07/24 18:13 & 11800 & C1.0 & 13/07/25 22:37 & 11800 & C2.1 \\
13/07/27 12:28 & 11800 & C1.0 & 13/07/28 19:57 & 11800 & C1.6 & 13/07/30 01:10 & 11801 & C1.1 \\
13/08/21 03:36 & 11820 & C1.3 & 13/08/22 19:08 & 11820 & C1.5$^{*}$ & 13/08/23 06:42 & 11827 & C1.1 \\
13/08/21 07:25 & 11828 & C2.2$^{*}$ & 13/08/22 20:23 & 11828 & C1.4 & 13/09/01 14:16 & 11834 & C1.7 \\
13/09/03 17:26 & 11834 & C1.3 & 13/08/30 02:04 & 11836 & C8.3$^{*}$ & 13/08/31 17:20 & 11836 & C2.6$^{*}$ \\
13/09/04 21:31 & 11836 & C1.3 & 13/09/05 19:46 & 11836 & C1.5 & 13/09/04 08:36 & 11837 & C2.6 \\
13/09/05 21:59 & 11837 & C1.4 & 13/09/24 22:50 & 11846 & C1.1$^{*}$ & 13/09/29 05:11 & 11850 & C1.7 \\
13/10/04 03:17 & 11856 & C2.5$^{*}$ & 13/10/07 15:34 & 11856 & C2.3 & 13/10/10 16:40 & 11861 & C1.3 \\
13/10/11 22:56 & 11861 & C6.3 & 13/10/12 22:02 & 11861 & C2.9 & 13/10/13 17:50 & 11861 & C4.5$^{*}$ \\
13/10/14 22:49 & 11861 & C7.4 & 13/10/15 17:04 & 11861 & C1.8$^{*}$ & 13/10/17 11:47 & 11861 & C4.8 \\
13/10/20 18:09 & 11873 & C1.7$^{*}$ & 13/11/08 18:16 & 11887 & C1.1 & 13/11/05 05:49 & 11889 & C1.6 \\
13/11/06 17:24 & 11889 & C3.0 & 13/11/14 08:55 & 11897 & C3.0 & 13/11/15 11:31 & 11897 & C7.5 \\
13/11/16 06:15 & 11897 & C8.6 & 13/11/17 21:16 & 11897 & C1.5 & 13/11/18 03:24 & 11897 & C2.8$^{*}$ \\
13/11/19 00:14 & 11897 & C1.8 & 13/11/21 09:03 & 11897 & C1.8 & 13/11/28 09:33 & 11907 & C1.0 \\
13/11/28 19:42 & 11908 & C1.0 & 13/12/12 03:11 & 11912 & C4.6$^{*}$ & 13/12/05 06:38 & 11916 & C1.9 \\
13/12/07 04:30 & 11916 & C3.3$^{*}$ & 13/12/11 22:49 & 11917 & C2.7 & 13/12/12 11:32 & 11917 & C2.3$^{*}$ \\
13/12/14 11:00 & 11917 & C2.3 & 13/12/16 08:37 & 11917 & C1.9 & 13/12/17 21:28 & 11917 & C1.9 \\
13/12/18 05:31 & 11917 & C1.5 & 13/12/12 22:05 & 11921 & C5.9 & 13/12/19 07:09 & 11930 & C1.8 \\
13/12/26 06:55 & 11931 & C2.2 & 14/01/03 18:30 & 11937 & C4.0$^{*}$ & 14/01/13 16:03 & 11952 & C1.8 \\
14/01/22 18:57 & 11955 & C1.3 & 14/02/09 05:12 & 11975 & C2.4 & 14/02/18 06:19 & 11976 & C1.9 \\
14/02/19 00:31 & 11976 & C1.2$^{*}$ & 14/03/02 15:47 & 11987 & C2.5 & 14/03/13 13:39 & 12003 & C1.7$^{*}$ \\
14/03/14 05:46 & 12003 & C4.7 & 14/03/19 11:19 & 12004 & C4.2$^{*}$ & 14/03/16 08:06 & 12005 & C2.2 \\
14/03/19 18:41 & 12010 & C2.2 & 14/03/25 08:05 & 12010 & C1.1 & 14/03/26 21:20 & 12010 & C1.2 \\
14/03/21 09:51 & 12013 & C2.7 & 14/04/04 13:34 & 12021 & C8.3$^{*}$ & 14/04/02 12:47 & 12026 & C1.2 \\
14/04/03 19:30 & 12026 & C1.0$^{*}$ & 14/04/04 03:43 & 12026 & C3.6 & 14/04/07 13:20 & 12026 & C1.7 \\
14/04/16 08:12 & 12034 & C4.5$^{*}$ & 14/04/19 05:55 & 12034 & C1.3 & 14/04/13 22:11 & 12036 & C1.0 \\
14/04/14 18:14 & 12036 & C1.0 & 14/04/15 13:22 & 12036 & C1.5 & 14/04/17 21:50 & 12036 & C3.2 \\
14/04/18 08:03 & 12036 & C4.8 & 14/04/19 12:47 & 12044 & C1.7$^{*}$ & 14/05/02 14:03 & 12047 & C1.4 \\
14/05/03 20:18 & 12047 & C1.8 & 14/04/28 15:17 & 12048 & C3.4$^{*}$ & 14/04/30 06:15 & 12049 & C1.5 \\
14/05/03 16:02 & 12049 & C1.7 & 14/05/04 15:54 & 12049 & C1.3 & 14/05/02 20:14 & 12052 & C2.1$^{*}$ \\
14/05/04 23:07 & 12053 & C1.1 & 14/05/05 11:57 & 12053 & C1.9 & 14/05/12 17:56 & 12060 & C1.1$^{*}$ \\
14/05/14 03:25 & 12060 & C1.6$^{*}$ & 14/05/15 14:02 & 12060 & C1.0 & 14/05/14 12:59 & 12063 & C8.3 \\
14/05/15 05:26 & 12063 & C3.2 & 14/05/16 20:11 & 12063 & C2.5$^{*}$ & 14/05/18 06:11 & 12063 & C3.8 \\
14/05/23 04:51 & 12065 & C1.5 & 14/05/24 13:30 & 12065 & C1.0 & 14/05/17 02:34 & 12066 & C3.4 \\
14/05/21 07:09 & 12066 & C1.6$^{*}$ & 14/05/22 13:13 & 12066 & C1.4$^{*}$ & 14/05/27 18:40 & 12071 & C1.1 \\
14/05/29 04:21 & 12071 & C1.4 & 14/06/11 02:56 & 12082 & C2.3$^{*}$ & 14/06/13 11:57 & 12082 & C2.5 \\
14/06/29 04:38 & 12096 & C1.1$^{*}$ & 14/06/28 08:22 & 12100 & C1.1 & 14/06/30 06:52 & 12100 & C2.2$^{*}$ \\
14/07/03 03:32 & 12100 & C2.6 & 14/07/01 10:04 & 12106 & C6.0 & 14/07/08 03:42 & 12106 & C1.2$^{*}$ \\
14/07/04 14:30 & 12108 & C4.2 & 14/07/05 08:28 & 12108 & C1.6 & 14/07/06 06:55 & 12108 & C3.5 \\
14/07/08 08:52 & 12108 & C4.1$^{*}$ & 14/07/04 05:34 & 12109 & C2.4$^{*}$ & 14/07/06 08:09 & 12109 & C2.9 \\
14/07/07 07:58 & 12109 & C4.3 & 14/07/09 04:05 & 12109 & C2.2 & 14/07/11 00:39 & 12109 & C4.6$^{*}$ \\
14/07/12 07:10 & 12109 & C2.5 & 14/07/13 09:50 & 12109 & C1.8 & 14/07/24 01:40 & 12121 & C2.1$^{*}$ \\
14/07/25 06:57 & 12121 & C2.2 & 14/07/28 13:56 & 12125 & C2.4 & 14/07/29 06:01 & 12126 & C1.8 \\
14/08/04 02:44 & 12134 & C2.1$^{*}$ & 14/10/04 06:48 & 12177 & C1.0 & 14/10/04 00:54 & 12178 & C1.7 \\
14/10/05 09:32 & 12178 & C1.0$^{*}$ & 14/10/06 15:44 & 12181 & C1.7 & 14/10/25 07:36 & 12195 & C9.2$^{*}$ \\
14/10/31 09:19 & 12201 & C2.0$^{*}$ & 14/11/01 18:14 & 12201 & C2.3 & 14/11/02 23:38 & 12201 & C1.9 \\
14/11/03 04:50 & 12201 & C1.4 & 14/11/08 18:53 & 12201 & C2.3 & 14/11/11 04:41 & 12208 & C3.4 \\
14/11/12 23:10 & 12208 & C1.5 & 14/11/13 06:54 & 12208 & C2.2 & 14/11/14 14:47 & 12208 & C3.1 \\
14/11/22 04:08 & 12216 & C1.7 & 14/11/24 22:08 & 12216 & C2.5 & 14/11/27 05:47 & 12216 & C2.8$^{*}$ \\
14/12/01 00:08 & 12216 & C4.2 & 14/11/26 06:09 & 12217 & C2.9$^{*}$ & 14/12/01 05:11 & 12217 & C1.8 \\
14/12/02 07:58 & 12217 & C5.2$^{*}$ & 14/12/03 02:30 & 12217 & C2.4 & 14/11/25 12:54 & 12219 & C1.1 \\
14/11/26 00:30 & 12219 & C1.7$^{*}$ & 14/11/28 20:12 & 12219 & C3.1 & 14/11/29 08:58 & 12219 & C2.1 \\
14/11/29 22:42 & 12221 & C1.7 & 14/11/30 17:16 & 12221 & C2.1 & 14/12/13 10:03 & 12227 & C4.0 \\
14/12/08 17:53 & 12230 & C1.3 & 14/12/09 23:22 & 12230 & C1.3 & 14/12/10 02:03 & 12230 & C1.4 \\
14/12/12 03:02 & 12230 & C1.1 & 14/12/10 17:07 & 12232 & C5.9 & 14/12/11 08:48 & 12234 & C3.8$^{*}$ \\
14/12/12 14:35 & 12234 & C4.0$^{*}$ & 14/12/21 04:47 & 12244 & C5.4$^{*}$ & 14/12/22 10:58 & 12244 & C7.0$^{*}$ \\
14/12/24 13:20 & 12245 & C2.0 & 14/12/26 08:54 & 12248 & C2.1$^{*}$ & 14/12/28 17:50 & 12248 & C3.3 \\
14/12/29 10:34 & 12248 & C1.4 & 14/12/29 10:46 & 12250 & C2.6$^{*}$ & 14/12/28 21:08 & 12251 & C1.3 \\
14/12/29 03:30 & 12251 & C1.4 & 14/12/30 06:10 & 12251 & C2.6$^{*}$ & 14/12/31 01:50 & 12251 & C1.6 \\
15/01/05 23:07 & 12251 & C4.5$^{*}$ & 15/01/04 04:16 & 12255 & C2.1 & 15/01/12 14:08 & 12255 & C7.1$^{*}$ \\
15/01/09 05:46 & 12259 & C3.4 & 15/01/10 04:58 & 12259 & C1.1 & 15/01/14 10:38 & 12259 & C1.9 \\
15/01/12 15:18 & 12260 & C3.7 & 15/01/11 23:29 & 12262 & C1.4$^{*}$ & 15/01/15 21:49 & 12262 & C1.2 \\
15/01/19 20:41 & 12266 & C3.3 & 15/01/27 05:43 & 12273 & C2.4 & 15/01/27 07:13 & 12275 & C2.1$^{*}$ \\
15/02/06 03:05 & 12281 & C1.1 & 15/02/09 07:04 & 12281 & C1.7 & 15/02/11 05:17 & 12282 & C1.0 \\
15/02/12 14:06 & 12282 & C1.0 & 15/02/18 21:53 & 12282 & C3.5$^{*}$ & 15/02/19 00:53 & 12282 & C1.2 \\
15/03/03 06:08 & 12292 & C1.9 & 15/03/04 10:12 & 12292 & C2.8 & 15/03/04 13:35 & 12293 & C1.1 \\
15/03/28 13:51 & 12303 & C1.2 & 15/03/25 13:38 & 12305 & C1.0 & 15/03/26 19:05 & 12305 & C1.4$^{*}$ \\
15/04/15 20:13 & 12321 & C7.9 & 15/04/16 16:16 & 12321 & C1.8 & 15/04/18 18:09 & 12321 & C2.9 \\
15/04/20 20:40 & 12321 & C2.4$^{*}$ & 15/04/23 07:31 & 12326 & C1.5 & 15/04/24 18:02 & 12331 & C1.0$^{*}$ \\
15/04/26 11:44 & 12331 & C1.0 & 15/05/04 02:49 & 12338 & C3.0 & 15/05/13 05:47 & 12342 & C3.5 \\
15/05/14 18:59 & 12342 & C1.1 & 15/05/15 22:18 & 12342 & C2.0$^{*}$ & 15/05/18 07:27 & 12349 & C1.0 \\
15/05/20 08:05 & 12349 & C1.6$^{*}$ & 15/05/21 06:57 & 12349 & C1.1$^{*}$ & 15/06/07 10:23 & 12362 & C1.6 \\
15/06/09 19:55 & 12364 & C2.8 & 15/06/10 00:10 & 12365 & C1.8 & 15/07/02 13:52 & 12373 & C1.1 \\
15/07/03 04:52 & 12373 & C1.2 & 15/07/10 16:07 & 12385 & C1.2 & 15/07/11 18:27 & 12385 & C1.0 \\
15/07/24 17:49 & 12389 & C2.6 & 15/08/07 18:11 & 12394 & C2.1 & 15/08/06 19:09 & 12396 & C2.1 \\
15/08/07 22:36 & 12396 & C1.7 & 15/08/08 14:17 & 12396 & C1.0 & 15/08/09 07:30 & 12396 & C4.2$^{*}$ \\
15/08/14 03:00 & 12401 & C1.6 & 15/08/15 12:02 & 12401 & C1.6 & 15/08/29 14:01 & 12405 & C1.4 \\
15/08/30 13:29 & 12405 & C1.0 & 15/09/11 21:30 & 12414 & C1.3 & 15/09/18 04:22 & 12418 & C2.6 \\
15/10/19 17:41 & 12436 & C1.3$^{*}$ & 15/10/21 17:48 & 12436 & C7.7 & 15/10/28 09:31 & 12436 & C1.9 \\
15/10/20 03:56 & 12437 & C1.2 & 15/10/26 10:21 & 12437 & C2.2$^{*}$ & 15/10/28 08:33 & 12437 & C1.6$^{*}$ \\
15/11/01 19:36 & 12441 & C1.8 & 15/11/02 09:50 & 12441 & C7.2 & 15/11/07 00:15 & 12448 & C3.0 \\
15/11/22 22:39 & 12454 & C2.7 & 15/11/23 01:09 & 12454 & C8.7$^{*}$ & 15/11/22 16:31 & 12457 & C2.0 \\
15/11/30 16:52 & 12458 & C1.0 & 15/12/01 07:57 & 12458 & C3.6 & 15/12/02 04:25 & 12458 & C2.2 \\
15/12/06 20:49 & 12463 & C1.1 & 15/12/07 22:39 & 12463 & C1.0$^{*}$ & 15/12/11 16:48 & 12465 & C5.6 \\
15/12/15 21:33 & 12468 & C1.2 & 15/12/16 08:34 & 12468 & C6.6$^{*}$ & 15/12/19 01:59 & 12468 & C3.1 \\
15/12/20 19:22 & 12468 & C2.4 & 15/12/17 12:22 & 12470 & C1.7$^{*}$ & 15/12/18 04:55 & 12470 & C4.6 \\
15/12/21 22:23 & 12470 & C1.1 & 15/12/23 04:02 & 12472 & C7.5 & 15/12/24 10:46 & 12472 & C2.0 \\
16/01/15 15:18 & 12480 & C1.7 & 16/01/20 14:25 & 12487 & C1.3 & 16/01/28 21:48 & 12488 & C3.3 \\
16/01/29 16:33 & 12488 & C1.2 & 16/01/27 13:26 & 12489 & C1.1 & 16/02/02 14:22 & 12491 & C1.2 \\
16/02/08 05:22 & 12492 & C1.6 & 16/02/12 06:27 & 12492 & C2.1 & 16/02/04 18:15 & 12494 & C5.2 \\
16/02/05 07:15 & 12494 & C2.9 & 16/02/06 22:37 & 12494 & C1.5$^{*}$ & 16/02/18 21:08 & 12501 & C1.8$^{*}$ \\
16/02/26 09:43 & 12506 & C1.0 & 16/02/27 05:44 & 12506 & C3.3 & 16/03/15 15:32 & 12521 & C1.0 \\
16/04/28 12:46 & 12535 & C1.9 & 16/05/01 09:14 & 12539 & C2.4 & 16/05/21 13:55 & 12546 & C1.0$^{*}$ \\
16/05/24 10:16 & 12546 & C1.3 & 16/05/26 13:45 & 12548 & C1.0 & 16/05/30 13:18 & 12550 & C1.0 \\
16/06/09 15:03 & 12552 & C1.1 & 16/07/07 07:49 & 12561 & C5.1 & 16/07/17 16:34 & 12565 & C1.0 \\
16/07/18 08:09 & 12565 & C4.4 & 16/07/19 10:00 & 12565 & C2.2 & 16/07/20 03:32 & 12565 & C2.5 \\
16/08/07 05:28 & 12571 & C1.3$^{*}$ & 16/08/08 20:13 & 12571 & C1.7$^{*}$ & 16/08/09 08:47 & 12574 & C2.5 \\
16/08/11 16:32 & 12574 & C2.4 & 16/08/28 21:25 & 12583 & C1.0 & 16/08/29 17:33 & 12583 & C1.1 \\
16/09/21 10:23 & 12593 & C1.6 & 16/09/27 07:33 & 12597 & C1.0$^{*}$ & 16/10/12 11:51 & 12599 & C1.1$^{*}$ \\
10/05/03 21:50 & 11066 & B1.0 & 10/05/05 16:11 & 11066 & B4.0 & 10/05/11 08:39 & 11068 & B1.1 \\
10/06/09 04:36 & 11078 & B5.1 & 10/07/01 20:54 & 11084 & B1.0 & 10/07/06 10:02 & 11086 & B1.0 \\
10/08/09 19:50 & 11095 & B1.6 & 10/08/10 08:48 & 11096 & B1.9 & 10/08/11 18:43 & 11098 & B1.7 \\
10/08/12 04:16 & 11098 & B1.4 & 10/08/16 02:20 & 11098 & B3.5 & 10/08/18 22:17 & 11100 & B3.3 \\
10/08/26 10:47 & 11101 & B2.6 & 10/08/28 00:44 & 11101 & B1.0 & 10/08/30 15:15 & 11102 & B1.4 \\
10/09/13 04:29 & 11106 & B2.1 & 10/09/15 22:20 & 11106 & B5.3 & 10/09/16 22:29 & 11106 & B4.6 \\
10/09/17 17:32 & 11106 & B2.7 & 10/09/18 06:18 & 11106 & B4.2 & 10/09/19 02:47 & 11106 & B2.4 \\
10/10/03 21:43 & 11111 & B1.7 & 10/10/05 08:51 & 11111 & B2.2 & 10/10/15 13:03 & 11113 & B1.0 \\
10/10/19 13:19 & 11113 & B5.1 & 10/10/23 17:29 & 11115 & B3.6 & 10/11/15 22:25 & 11126 & B8.3 \\
10/11/16 07:10 & 11126 & B3.6 & 10/11/17 04:35 & 11126 & B7.8 & 10/11/18 13:03 & 11126 & B2.4 \\
10/11/23 08:59 & 11127 & B1.3 & 10/11/25 20:59 & 11127 & B1.2 & 10/11/25 00:46 & 11128 & B1.4 \\
10/12/03 06:51 & 11131 & B5.3 & 10/12/05 13:02 & 11131 & B2.2 & 10/12/08 05:05 & 11131 & B1.4 \\
10/12/09 03:17 & 11131 & B1.3 & 10/12/11 11:14 & 11131 & B4.0 & 10/12/04 21:11 & 11132 & B5.6 \\
10/12/05 02:05 & 11132 & B2.5 & 11/01/01 21:52 & 11140 & B8.3 & 11/01/11 06:06 & 11146 & B1.4 \\
11/01/29 16:27 & 11150 & B1.8 & 11/02/04 19:30 & 11152 & B6.9 & 11/02/10 20:49 & 11156 & B5.3 \\
11/02/09 19:57 & 11157 & B2.9 & 11/03/18 11:40 & 11173 & B3.3 & 11/04/04 03:07 & 11180 & B8.6 \\
11/04/29 12:44 & 11200 & B6.6 & 11/05/03 06:20 & 11200 & B6.3 & 11/04/28 09:15 & 11202 & B6.7 \\
11/05/16 22:11 & 11214 & B3.5 & 11/05/17 01:04 & 11214 & B3.5 & 11/05/25 22:25 & 11223 & B3.0 \\
11/05/27 09:50 & 11223 & B4.4 & 11/06/01 05:01 & 11229 & B7.3 & 11/06/29 09:47 & 11242 & B5.5 \\
11/07/07 15:49 & 11245 & B3.1 & 11/07/20 17:51 & 11259 & B5.1 & 11/08/27 04:42 & 11275 & B8.7 \\
11/09/10 17:16 & 11291 & B8.1 & 12/04/08 23:34 & 11451 & B4.1 & 12/05/26 21:26 & 11490 & B5.4 \\
12/05/28 14:09 & 11490 & B6.8 & 12/12/10 03:32 & 11630 & B4.7 & 12/12/11 03:02 & 11630 & B5.6 \\
12/12/12 00:02 & 11630 & B5.7 & 13/02/18 04:23 & 11673 & B4.0 & 13/02/28 08:24 & 11680 & B4.7 \\
13/02/27 03:25 & 11682 & B8.2 & 13/03/01 10:11 & 11682 & B6.8 & 13/03/25 17:43 & 11704 & B4.4 \\
13/05/30 12:58 & 11757 & B9.5 & 13/06/13 21:06 & 11768 & B3.6 & 13/08/02 11:08 & 11806 & B9.7 \\
13/08/02 07:57 & 11807 & B5.7 & 13/07/30 20:59 & 11808 & B8.7 & 13/08/03 05:48 & 11810 & B3.6 \\
13/08/07 14:42 & 11810 & B4.7 & 13/09/08 16:29 & 11838 & B4.1 & 13/09/18 09:51 & 11847 & B4.1 \\
13/10/06 00:16 & 11857 & B8.7 & 15/02/15 00:46 & 12283 & B9.0 & 15/04/07 03:32 & 12318 & B8.2 \\
15/05/20 15:02 & 12351 & B5.8 & 15/07/10 01:21 & 12384 & B7.3 & 15/07/17 15:53 & 12387 & B5.1 \\
15/08/03 05:37 & 12391 & B5.3 & 15/08/01 01:14 & 12393 & B8.0 & 15/08/13 10:30 & 12400 & B5.8 \\
15/09/04 05:13 & 12409 & B4.4 & 15/09/05 10:06 & 12409 & B2.2 & 15/09/18 14:34 & 12419 & B6.0 \\
15/10/14 15:46 & 12432 & B7.6 & 15/11/29 06:55 & 12459 & B6.2 & 16/01/03 20:20 & 12476 & B4.7 \\
16/01/25 21:54 & 12490 & B4.7 & 16/02/07 06:55 & 12495 & B7.0 & 16/02/21 13:00 & 12505 & B5.1 \\
16/03/06 04:58 & 12512 & B7.2 & 16/03/18 23:03 & 12519 & B6.7 & 16/03/25 14:55 & 12526 & B6.5 \\
16/03/26 07:19 & 12526 & B1.6 & 16/03/27 10:28 & 12526 & B1.7 & 16/03/28 07:52 & 12526 & B3.7 \\
16/04/26 11:37 & 12536 & B5.7 & 16/04/27 03:47 & 12536 & B7.2 & 16/04/30 12:10 & 12536 & B5.2 \\
16/06/16 00:20 & 12553 & B3.2 & 16/06/16 15:17 & 12555 & B3.8 & 16/06/17 14:00 & 12555 & B1.9 \\
16/07/13 14:40 & 12562 & B4.2 & 16/08/17 09:56 & 12576 & B3.5 & 16/08/16 08:08 & 12578 & B8.1 \\
16/08/20 05:51 & 12578 & B4.4 & 16/08/26 06:50 & 12581 & B6.6 & 16/09/17 08:08 & 12592 & B1.9 \\
16/09/18 10:41 & 12592 & B6.8 & 16/10/02 18:53 & 12598 & B3.4 & 16/10/05 01:10 & 12598 & B5.6 \\
16/10/16 13:29 & 12602 & B3.5 & 16/11/13 17:29 & 12610 & B1.5 & 16/11/14 23:37 & 12610 & B1.1 \\
16/11/15 16:53 & 12610 & B4.2 & 16/11/19 07:10 & 12611 & B1.9 & 16/11/30 05:22 & 12614 & B3.9 \\
16/12/11 19:16 & 12617 & B1.1 & 16/12/27 10:04 & 12621 & B1.6 &  &   &   \\
\enddata
\tablecomments{These listed 845 samples include 23 X-class, 142 M-class, 552 C-class, and 128 B-class flaring ARs. The 142 C-class ARs marked with a $^*$ are randomly selected from the 552 C-class AR samples (see Section~\ref{method}); together with AR samples in other classes, they form a data set for use in Section~\ref{features}.}
\end{deluxetable*}

\clearpage
\begin{figure}[!t]
\epsscale{.47}
\plotone{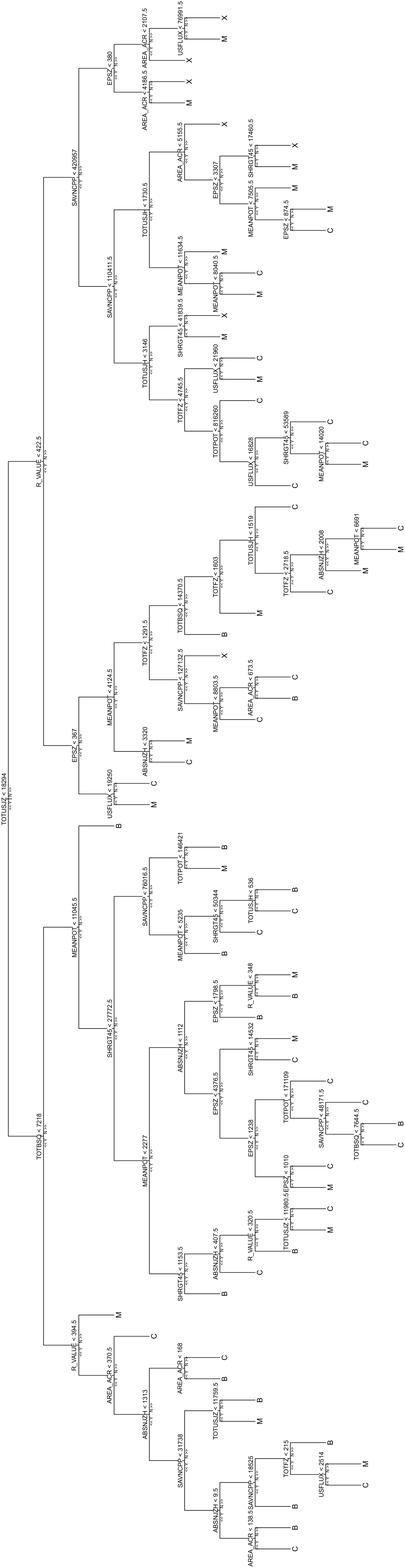}
\caption{A sample representative tree grown by the RF algorithm. \label{tree} }
\end{figure}

\end{document}